# Low thermal boundary resistance at bonded GaN/diamond interface by controlling ultrathin heterogeneous amorphous layer

Bin Xu[1,2#], Fengwen Mu[3#*], Yingzhou Liu[4], Rulei Guo[1], Shiqian Hu[4**], Junichiro Shiomi[1,2***]

[1] Department of Mechanical Engineering, The University of Tokyo, 7-3-1 Hongo, Bunkyo, Tokyo 113-8656, Japan.

[2] Institute of Engineering Innovation, The University of Tokyo, 2-11 Yayoi, Bunkyo, Tokyo 113-8656, Japan.

[3] Innovative Semiconductor Substrate Technology Co., Ltd. No.25, Huayuan North Road, Haidian District, Beijing 100083, China.

[4] School of Physics and Astronomy, Yunnan University, 650091 Kunming, China.

[#]These authors contribute equally

Corresponding author:

*mufengwen@isaber-s.com

**shiqian@ynu.edu.cn

***shiomi@photon.t.u-tokyo.ac.jp

## Abstract

Thermal boundary resistance (TBR) in semiconductor-on-diamond structure bottlenecks efficient heat dissipation in electronic devices. In this study, to reduce the TBR between GaN and diamond, surface-activated bonding with a hybrid SiOx-Ar ion source was applied to achieve an ultrathin interfacial layer. The simultaneous surface activation and slow deposition of the SiOx binder layer enabled precise control over layer thickness (2.5–5.3 nm) and formation of an amorphous heterogeneous nanostructure comprising a SiOx region between two inter-diffusion regions. Crucially, the 2.5-nm-thick interfacial layer achieved a TBR of 8.3 m2·K/GW, a record low for direct-bonded GaN/diamond interface. A remarkable feature is that the TBR is extremely sensitive to the interfacial thickness; rapidly increasing to 34 m2·K/GW on doubling the thickness to 5.3 nm. Theoretical analysis revealed the origin of this increase: a diamond/SiOx interdiffusion layer extend the vibrational frequency, far exceeding that of crystalline diamond, which increases the lattice vibrational mismatch and suppresses phonon transmission.

## Keywords

Thermal boundary resistance, vibrational mismatch, GaN, diamond, amorphous interface, power device, heat dissipation

# Introduction

Recently, the development of electronic devices based on wide-bandgap semiconductors such as gallium nitride (GaN) and silicon carbide (SiC) has enabled the commercialization of electric vehicles, mobile devices, and 5G-and-beyond telecommunication[1,2]. However, increases in device power and size miniaturization, cause problems with device cooling because of the increased self-heating[3], which can degrade performance[4]. Integrating the device on high thermal conductive diamond substrate is a promising solution to enhance cooling efficiency. This strategy is also highly valuable for other electronic devices such as those based on silicon[5] and indium phosphide[6]. However, although the high thermal conductivity ($\kappa \approx 2000$ W/m·K) of diamond can enhance heat dissipation, the semiconductor/diamond interface introduces thermal resistance in the heat dissipation pathway, the so-called thermal boundary resistance (TBR). Therefore, interfacial structure design and integration methods[7,8] are key to reduce the TBR and maximize heat dissipation.

The large difference in the Debye temperature between diamond and SiC, GaN, or Si shows that the semiconductor-on-diamond interface is mismatched, having distinctly different phonon properties on each side. Commonly used empirical mismatch models, such as the acoustic mismatch model (AMM) and diffuse mismatch model (DMM), cannot be applied to predict the TBR because they only consider the bulk phonon states. Increasingly, studies have shown how interfacial structures and states can influence the TBR, sometimes counterintuitively. For instance, Xu et al. found that when bonding copper and diamond with a self-assembled monolayer (SAM), the TBR decreases when copper and the SAM are bonded by weak van der Waals interactions rather than strong covalent bonds[9]. In addition, Yang et al. reported that thicker amorphous carbon interfacial layers can result in lower TBRs at the copper/diamond interface[10].

In the semiconductor-on-diamond structure, a buffer layer such as Si or silicon oxide ($SiO_2$) is required because of the lattice constant differences, making direct bonding challenging[3]. For example, for surface-activated bonding (SAB) at the GaN/diamond interface, the Ga-enriched GaN surface resulting from Ar-ion bombardment cannot directly bond with amorphous carbon layer on the ion-bombarded diamond surface without a binder layer. Moreover, to prevent thermal stress and damage[11], a low-temperature or room-temperature process is required for high-quality semiconductor-on-diamond structures, resulting in an amorphous interfacial layer. Such amorphous structures are frequently formed when interfacial growth involves nonepitaxial deposition via sputtering, atomic layer deposition (ALD), or vacuum evaporation processes. Crucially, the heat conduction of amorphous materials is different from those of crystalline materials. Two types of delocalized heat carriers, propagons and diffusons, are responsible for heat conduction. In addition, the localized heat carriers, locons, can also influence the interfacial heat conduction by providing localized interfacial states[12,13]. Although difficulties in experimental measurements and nanostructure fabrication have limited research on heat conduction at amorphous interfaces[13], computational studies have identified the anomalous behavior of amorphous interfaces. For example, Giri et al.[14] and Gordiz et al.[15] found that the TBR at the amorphous Si/Ge interface could be as low as 0.5 m²·K/GW, far below that of its crystalline counterpart; notably, this is inconsistent with the expectation that disorder leads to higher thermal resistance. Further, Giri et al. found that increasing disorder enhances the vibrational density of states (vDOS) of interfacial modes, reducing the TBR at the hydrogenated amorphous silicon carbide/hydrogenated amorphous silicon oxycarbide interface[16]. For semiconductor-on-diamond systems, the TBR of GaN/diamond interfaces having different amorphous interfacial layer thicknesses fabricated using conventional SAB, where Si is used as the binder layer, have been investigated[11,17,18], and the near linear dependence of the TBR on interfacial layer thickness has been observed, suggesting the diffusive transport of heat carriers in the Si interfacial layer. Moreover, because the studied interfacial layer thicknesses ranged from 5 to 25 nm, the effective mean free path of the heat carrier must be smaller than 5 nm. Therefore, the TBR at small interfacial layer thicknesses should be studied. However, the precision fabrication of ultrathin interfacial layers remains challenging.

In this study, we investigated the TBR of a GaN/diamond system, a conventional semiconductor-on-diamond system. To achieve this, a room-temperature SAB technique employing a hybrid ion (Ar and $SiO_x$) process was used, enabling the fabrication of ultrathin interfacial layers and precise thickness control. The interfacial layer includes a $SiO_x$ intermediate region between two diffusive regions resulting from the diffusion of $SiO_x$ into the surface of GaN and diamond on the individual sides of the interface. Notably, a record low TBR (8.3 $m^2 \cdot K/GW$) was obtained at an interfacial thickness of 2.5 nm. In addition, the TBR was very sensitive to thickness: on increasing the thickness to 5.3 nm, the TBR increased to 34 $m^2 \cdot K/GW$. Molecular dynamics (MD) simulations revealed that the increase in TBR results from the growth of the interdiffusion layer, which increases the frequency range of the vDOS. In particular, at the $SiO_x$–diamond interface, the frequency reached 75 THz, nearly 1.7 times that of crystalline diamond, resulting in a significant mismatch at the interface and transmission suppression.

## Results
### Sample preparation and characterization

To fabricate GaN on the diamond sample, room-temperature bonding instead of the chemical vapor deposition (CVD) growth was used because CVD cannot control the interfacial structure precisely and causes thermal damage and stress. Specifically, an SAB method using a hybrid Ar and $SiO_x$ ion source was developed, denoted as "hybrid SAB." Importantly, the hybrid ion source and the use of $SiO_x$ ion enabled simultaneous surface activation and binder layer deposition with slow rate, resulting in a thin binder layer and control over the interfacial structure. First, GaN and diamond wafers were surface-treated using $Ar/SiO_x$ hybrid ion sources in a high-vacuum chamber to create an activated binder layer comprising amorphous $SiO_x$ (Fig. 1(a)). This process assists bonding when the activated surfaces are pressed together at room temperature inside the vacuum chamber (Fig. 1(b)). Using this precise hybrid SAB method, four samples (#1–#4) having thin interfacial layers ranging from 2.5 to 5.3 nm were fabricated, as confirmed by scanning transmission electron microscopy (STEM) measurements.

STEM images of the interfacial layer and energy-dispersive X-ray spectroscopy (EDS) elemental line scans near the interfacial layer of sample #1 are shown in Fig. 1(d) and 1(i), respectively (see Figs. 1(f)–(k) for full EDS maps). The results for samples #2–#4 are shown in Figs. S1, S2, and S3. The STEM image suggests that the interfacial layer is amorphous (Fig. 1(d)) having a heterogeneous layered nanostructure and comprising three regions: diffusive region 1 (mainly C, Si, and O), $SiO_x$, and diffusive region 2 (mainly Ga, N, Si, and O), from the GaN to the diamond sides (Fig. 1(e)). These two diffusive regions were generated during the initial depostition/activation stages, where accelerated Si and O ions diffused into the amorphized surface layer of the GaN or diamond, which was simultaneously bombarded with Ar ions. The $SiO_x$ region was generated in the later stages of the process and is located far from the amorphized GaN and diamond surfaces. As shown by STEM measurements, there were no sharp interfaces between regions because of the nature of interfacial diffusion. Such heterogeneous layered amorphous nanostructures are inevitable at the amorphous SAB interface, where accelerated high-energy ions diffuse into the surface of the wafer during the sputtering process.

To facilitate discussion of the interfacial structure and TBR, we characterized the thickness of the constituent layers as follows: using EDS line scans, the point at which the Si intensity decreased to half its maximum value was defined as the boundary of the $SiO_x$ region. The thickness of the diffusive region was calculated by subtracting the thickness of the $SiO_x$ region from the total interfacial thickness. The total thickness was obtained from STEM imaging, wherein the boundary was defined as the location where the crystal region sharply transits into the amorphous region. The thickness of the diffusive region was similar on both sides of all samples. The dimensions of the interfacial layers of samples #1–#4 are listed in Table 1. For samples #1–#3, the amorphous $SiO_x$ layers ranged from 1.9–2.0 nm, and the difference in the total interface thickness originates mainly from the diffusion layer. For sample #4, a thicker $SiO_x$ layer (3.1 nm) was obtained, but the thickness of the diffusive layer was equal to that of sample #3.

### Low thermal boundary resistance at GaN/diamond interface

Next, we investigated the TBR of the GaN/diamond interfaces ($R_{GaN-Dia}$) using the time-domain thermoreflectance (TDTR) method. A schematic of the three-layer model and the TDTR measurement are shown in Fig. 1(c). To obtain $R_{GaN-Dia}$, the TBR of the Al/GaN interface ($R_{Al-GaN}$) and the thermal conductivity of GaN ($\kappa_{GaN}$) must be known. This was achieved by changing the modulation frequency to 11.05 MHz, such that the penetration depth of laser-induced heating decreased to one-third of that at 1.111 MHz. This allowed us to measure $R_{Al-GaN}$ and $\kappa_{GaN}$ independently. Calculations revealed that the sensitivity of $R_{Al-GaN}$ and $\kappa_{GaN}$ can reach 0.6 and 0.3, respectively, significantly higher than that of $R_{GaN-Dia}$ (Fig. S4). Fig. 2(a) shows the $R_{Al-GaN}$ and $\kappa_{GaN}$ of samples #1–4. The difference in $R_{Al-GaN}$ is small, mainly originating from the slight roughness of the GaN surface as a result of the polishing process. The value for $R_{Al-GaN}$ was similar to that obtained in our previous study using an identical polishing process[11]. The variance in $\kappa_{GaN}$ across samples #1–4 can be attributed to the disparities in the thickness of the GaN layers. This is due to the fact that their thickness falls below the maximum mean free path of phonons in GaN, which significantly influences thermal conductivity. The longest phonon MPF in GaN, which has a substantial impact on thermal conductivity, is known to extend up to 3 μm[19]. The κ GaN value is consistent with previous results for similar GaN thicknesses[18]. Subsequently, we changed the modulation frequency to 1.111 MHz to increase the penetration depth for $R_{GaN-Dia}$. As suggested by sensitivity calculations, the sensitivity of RGaN-Dia approached 0.5, enabling the extraction of precise RGaN-Dia from the temperature decay profile using the pre-obtained $R_{Al-GaN}$ and $\kappa_{GaN}$ (Fig.2(b)). $R_{GaN-Dia}$ increased from 8.3 to 35.0 m²·K/GW on increasing the interface from 2.5 to 5.3 nm (Fig. 2(c), inset). The quadrupling of the TBR on doubling the thickness demonstrates the sensitivity of the TBR to changes in interfacial structure (see Sections 4 and 5). Moreover, the TBR of 8.3 m²·K/GW is a record low in direct bonded GaN/diamond interface, which may impact the efficient heat dissipation of GaN device.

**Finite element method (FEM) analysis of heat dissipation in GaN/diamond devices**
To investigate the impact of the low $R_{GaN-Dia}$ on heat dissipation, we conducted FEM analysis by solving Fourier's law of heat conduction using COMSOL Multiphysics 5.3a to study the temperature of high-power GaN devices. Fig. 3(a) shows a schematic of the GaN-on-diamond device mounted on a water-cooling plate. The simulations parameters, per previous studies, are listed in Table S1[20,21]. The temperature profile obtained by FEM simulations is presented in Fig. 3(b). As shown, the device temperature was 128 °C when $R_{GaN-Dia}$ and $\kappa_{diamond}$ were 8.3 m²·K/GW and 2000 W/m·K, respectively (i.e., the experimental values). We also examined cases using $R_{GaN-Dia}$ and $\kappa_{diamond}$ reported previously (Fig. 3(c)), focusing on studies integrating GaN with diamond via CVD[1,22,23] and bonding methods[18,24,25]. Compared with the bonding method, direct growth involves simultaneous interface formation and material synthesis, reducing controllability over the interfacial structure and diamond crystallinity and, consequently, resulting in a high TBR or lattice thermal conductivity. Literature values of $R_{GaN-diamond}$ and $\kappa_{diamond}$ are listed in Table S2 (see experimental for FEM parameters). For the bonding method, $\kappa_{diamond}$ was similar, and we observed a nearly linear correlation between device temperature and $R_{GaN-Dia}$. Specifically, the device temperature is approximately 1.5 °C lower than the previously reported lowest $R_{GaN-Dia}$ for a bonded GaN/diamond interface[18]. Concerning the simulation using the parameters from the synthetic method, there is a tradeoff between the $R_{GaN-Dia}$ and the thermal conductivity of diamond, balanced by the growth parameters.

As a results, although the literature $R_{GaN-Dia}$ is lower than that of the present work1, the reduced thermal conductivity of the CVD synthesized diamond layer significantly sacrifices the heat dissipation efficiency, resulting in a much higher device temperature of 157.5 °C, approximately 30 °C higher than that reported in the present study. The FEM simulations suggest that the SAB bonding method is better for achieving highly efficient heat dissipation, particularly when a low TBR is achieved.

**Dependence of thermal resistance on interfacial layer thickness**
Fig. 2(c) shows the dependence of $R_{GaN-Dia}$ on interfacial layer thickness (L), as well as literature values for amorphous interfacial layers fabricated using conventional SAB[18,24]. As shown, the L-dependence in the current work is remarkably different; in particular, the slope is much larger, and the trend is nonlinear. Further, the small difference in L of 2.2 nm between samples #2 and #4 resulted in a large difference in $R_{GaN-Dia}$ up to 23.8 m2·K/GW. Assuming the additional thermal resistance with increase in L is due to the thermal resistance of the interfacial layer itself, the equivalent thermal resistivity would be 10.8 m·K/W, i.e., 0.09 W/m·K in terms of effective thermal conductivity. This is significantly smaller than 0.79 W/m·K, the literature effective thermal conductivity of the interfacial layer extracted from linear fitting of $R_{GaN-Dia}$[18,24] and 1.3–1.5 W/m·K, the typical thermal conductivity of amorphous $SiO_2$[26]. Note that the thermal conductivity of amorphous $SiO_2$ should not decrease significantly with a decrease in thickness[27] because the propagon contribution is small (approximately 6 %[26]). In addition, although the diffusive region is not pure amorphou $SiO_x$ (a-$SiO_x$), because propagons are not the predominant heat carriers, their scattering by impurities such as Ga, N, or C has a negligible influence on overall thermal conductivity. Even in amorphous carbon, where propagons significantly contribute to thermal conductivity, impurities only reduce the thermal conductivity from 2.6 to 1.5 W/m·K[28]. Impurities also have other effects than scattering propagons, as shown by studies of $SiO_2$ deposited with impurities such as Ar and Fe. Although the thermal conductivity can be suppressed to 0.71 W/m·K[16,29–32], the reduction is much smaller than that obtained in this study. Thus, factors other than the thermal resistance of the amorphous interfacial layers are responsible for the large $R_{GaN-Dia}$, while the heat carriers behave more complex than the diffusion behavior in previous study.

Regarding interfacial thermal conductance, the overlap of the vDOS at the interface is also critical. As reported previously, the vDOS distributions of GaN and diamond differ, as indicated by their cutoff frequencies: 30 and 50 THz, respectively[24]. Crucially, the vibrational frequency overlap near the interface determines the TBR. Thus, the overlap of the vDOS at the diamond and GaN interfacial layer, known as the "bridging effect," is important. Therefore, next, we investigated the vDOS at the interface.

**MD analysis for interfacial thermal conduction**
To explore the mechanism behind the unexpected TBR variation, we conducted MD simulations. First, a heterogeneous layered amorphous nanostructure was built through step-by-step annealing (Fig. 4(a)), yielding the experimentally observed GaN/diamond interface. The system consisted of a $SiO_x$ region ($L_{a-SiOx}$) with a thickness of either 1.9 or 2.95 nm and two diffusive regions with individual thickness ranging from 0.5 to 4 nm. At a fixed $L_{a-SiOx}$ of 1.9 nm, with the increase in L, we observed an increase in $R_{GaN-Dia}$, and the rate of growth increased at larger L (Fig. 4(b)), consistent with the experimental results shown in Fig. 2(c). Moreover, the TBR of the heterogeneous multilayer interface is larger than that of its homogeneous amorphous counterpart consisting of pure SiO2 (Fig. S5), again consistent with the experimental observation that the heterogeneous amorphous interfacial layer provides a larger thermal resistance.

Next, the vDOS was calculated to examine the frequency overlap at the interface. First, we considered the entire interfacial layer. As shown in Fig. 4(d), the vDOS of the entire interfacial region was similar, regardless of thickness (Fig. 4(d)). We focused on the vDOS overlap between the GaN/interfacial layer and diamond/interfacial layer to assess the bridging of the vibrational mismatch. However, the vDOS overlap was comparable, and there was no distinct correlation with the TBR, regardless of interfacial thickness. Thus, considering vibrational bridging over the entire interfacial layer cannot explain the observed phenomena, and examination at different locations within the interfacial layer is required.

Therefore, we calculated the local vDOS along the cross-interface direction of the interfacial layer (Figs. 4(f)–(h)), revealing distinctive local vDOS in different interfacial regions despite the small thickness and blurt interface in between. The $SiO_x$ region exhibits lower-frequency vibrational states than the two diffusive regions. Meanwhile, the diamond-$SiO_x$ diffusive region has a significantly extended vDOS at 75 THz. These high-frequency vibrations could result from the amorphous carbon phase in this region, which typically has a much higher frequency than its crystal counterpart[33] because of stronger bonding[34]. Moreover, compared with the relatively small variation in the vDOS of the $SiO_x$ and GaN-$SiO_x$ regions, apparent variations appear in the diamond-$SiO_x$ diffusive region: specifically, the high-frequency component of the vDOS increases, whereas the low-frequency component decreases with the increase in L (Fig. 4(h)). This variation in the vDOS of the $SiO_x$/diamond layer can be attributed to the variations in the percentages of $SiO_x$ and amorphous carbon resulting from $SiO_x$ diffusion into the diamond region. Further, we extracted the elemental concentrations in the SiOx/diamond layer, as shown in Fig. S6. With increase in diffusive layer thickness, the carbon concentration increased, whereas those of Si and O decreased. Because the vDOS of $SiO_x$ and diamond are significantly different, the concentration variations have a significant effect on the overall vDOS, where a thicker layer with fewer $SiO_x$ components can give rise to more high-frequency vibrations. Fig. S7 shows a schematic of the cutoff frequency along the through-plane direction. Compared to a previous study that used a-Si as a binder layer, the cutoff frequency of a-$SiO_x$ is 55 THz, which is much higher than the value (18 THz) of amorphous Si[35]. Moreover, the two diffusive layers having even higher cutoff frequencies can enhance the vDOS mismatch in the heterogeneous nanostructured interfacial layer.

To further study the influence of the variation in the vDOS of the $SiO_x$/diamond layer, we analyzed the heat-carrier transmission at identical locations (marked in Fig. 4(a)). The overall heat-carrier transmission decreased with the increase in the thickness of the diffusive region, consistent with the TBR calculations. Notably, a significant decrease in the transmission of heat carriers occurred within the low-frequency regime below 30 THz, whereas the high-frequency regime starting at 30 THz remained relatively unchanged (Fig. 4(c)). This result agrees with the suppressed vDOS within 0–30 THz in the diamond-$SiO_x$ diffusive region, which results in a poor overlap of the vibrational frequency between the crystal diamond and the a-SiO$_x$ region (Fig 4(i)). The analysis indicates that the local vDOS is critical for determining the interfacial heat conduction and is responsible for the significantly enhanced TBR at larger L.

**Summary**

To enhance heat dissipation for semiconductor-on-diamond systems, we conducted a comprehensive study of the interfacial heat conduction in a GaN-on-diamond system. Using the hybrid SAB method with Ar/SiO$_x$ ion source, we fabricated ultrathin interfacial layers with precisely controlled nanostructures. This interfacial layer has a heterogeneous layered structure comprising a SiO$_x$ region between two diffusive regions. By reducing the total thickness of the interfacial layer, we achieved a record low TBR of 8.3 m2·K/GW at the bonded GaN/diamond interface. This low TBR could enhance the performance of actual GaN devices, as shown by FEM analysis. However, small changes in the thickness of the individual layers can significantly affect the TBR. Notably, the simulations clarified the impact of the significantly extended vDOS of the interfacial layer, particularly in the SiO$_x$/diamond region, whose cutoff frequency was nearly 1.7 times that of crystal diamond. The vDOS is also highly sensitive to the interfacial layer thickness; that is, a thinner layer can lead to an increase in the vDOS in the high-frequency regime and a decrease in the low-frequency regime. This variation can result in a large interfacial vibrational frequency mismatch and significantly impede the transmission of low-frequency heat carriers. The findings of this study not only reveal that the hybrid SAB method using Ar and SiO$_x$ ion sources is a promising method for wafer bonding with thin and precisely controlled interfacial layers for low TBR, but also stress the importance of controlling the amorphous nanostructure of the thin interfacial layer for achieving a low TBR. The methodology established in this study can address the practical challenge in enhancing the efficiency of heat dissipation in high-performance electronic systems. Moreover, the advanced fundamental understanding on the interfacial heat conduction across amorphous interfacial layer provides crucial insight into the design of the interface in the next-generation thermal management in semiconductor devices.

## Method
### Sample Preparation
Single-crystal diamond (EDP Co., Ltd) was fabricated using CVD in dimensions of 10 mm × 10 mm and 500-μm thickness and bonded to a GaN (2 μm)/sapphire (430 μm) substrate (Powerdec Co., Ltd) using a modified room-temperature SAB method. Initially, chemical–mechanical polishing was performed on the GaN and diamond surfaces to reduce the root-mean-square surface roughness to less than 0.5 nm. Unlike the conventional SAB method, where the binder layer is sputtered on the surface of GaN and diamond before surface activation, we used a $SiO_x$-Ar beam source, enabling simultaneous deposition of the $SiO_x$ binder layer and surface activation. This method enables the generation of an ultrathin binder layer with precise control over the interfacial thickness. The acceleration voltage and current of the ion-beam source were 100 mA and 1.0 kV, respectively. Details of the SAB process and the differences between standard and hybrid SAB can be found in previous studies[11,36]. After the SAB process, the sapphire substrate was removed via laser lift-off, followed by the evaporation of the aluminum layer, which acted as a transducer in the TDTR measurement.

### Time-Domain Thermal Reflectance
The TDTR method was used to investigate the TBR between GaN and diamond ($R_{GaN-diamond}$). The setup details and operating principles of the TDTR were identical to those used previously[9,37]. Briefly, TDTR is a typical pump–probe method for studying nanoscale heat conduction by measuring the pump-induced temperature response at the top of a sample using a probe laser[38,39]. In the TDTR setup, a pulsed Ti:sapphire laser (approximately 800 nm) with a repetition rate of 80 MHz and a pulse width of 140 fs was used. The laser was split into two paths: pump and probe. The pump laser was converted from 800 to 400 nm using a nonlinear crystal, followed by frequency modulation using an electro-optic modulator. The delay time of the probe laser was controlled by varying the light path of the delayer stage. The spot sizes of the pump and probe beams were 17.5 and 10.1 μm, respectively. A Si photodiode was used to detect the probe laser, the signal of which was monitored using a connected lock-in amplifier at a defined modulation frequency. We considered the ratio $(V_{in}(t)/V_{out}(t))$[40,41] of the detected in-phase voltage ($V_{in}$) and out-of-phase voltage ($V_{out}$) as the temperature response.

Moreover, the modulation frequency is closely correlated with the pump-laser-induced heating penetration depth, enabling us to measure the targeted thermal properties in the heat conduction model at different depths from the surface selectively. Frequencies of 1.111 and 11.05 MHz were applied to measure $k_{GaN}$ and $R_{GaN-diamond}$ separately. The sensitivity calculation results are shown in Fig. S4, and the parameters used in the TDTR analysis (sensitivity calculation and fitting) are listed in Table S3. Following conventional TDTR data processing, we used the thermal boundary conductance (TBC) values for the aluminum/GaN and GaN/diamond interfaces ($G_{Al-GaN}$ and $G_{GaN-diamond}$, respectively). To fit the individual samples, the thickness of each layer was directly obtained from the STEM images shown in Figs. S8 and S9. The TBR and thermal conductivity were obtained following the procedure described in the main text.

### Material Characterization
A focused ion beam system (Thermo Scientific Helios G4 HX) with dual beams was used to prepare cross-sectional transmission electron microscopy (TEM) samples having thicknesses of approximately 50 nm for analysis of the interface. For STEM measurements, we used a Thermo Scientific Themis Z spherical aberration-corrected transmission electron microscope operating at an accelerating voltage of 200 kV in bright field (BF) and high-angle annular dark-field (HAADF) modes. For TEM measurements, we used a Thermo Scientific Tecnai F20 transmission electron microscope operating at an acceleration voltage of 200 kV in bright-field (BF) TEM mode. In HR-STEM mode, the spatial distribution of elements was obtained using EDS mapping. In this study, we used the mold density signal in the EDS mapping and line-scan data. To determine the dimensions of the interfacial structures, the STEM images were used to define the total thickness of the interfacial layer. For the thickness of the $SiO_x$ region, we used EDS results. Because the spatial resolution of EDS mapping is approximately 0.5 nm, there is a small systematic error in the $SiO_x$ layer thickness, as presented in Table 1. Further, we not only measured the GaN/diamond interfacial region to analyze the detailed interfacial structure but also measured the GaN layer (Fig. S8) and an aluminum layer (Fig. S9) to obtain precise values for TDTR analysis.

**MD Simulations**

The TBR at the interface between GaN and diamond was investigated using nonequilibrium molecular dynamics (NEMD) with the LAMMPS package[42]. The interatomic interactions in crystalline diamond[43,44], GaN[45], and $SiO_2$[46] were modeled using established Tersoff potentials. For interatomic interactions involving different atoms within the amorphous region, specific rules within the framework of the Tersoff potential were followed[45–48] (see SI Part 2). The simulations model for the GaN-on-diamond structure with fixed system dimensions of 23 nm × 4.5 nm × 4.3 nm is shown in Fig. 3(a). The central part of the system consists of an amorphous region of amorphous (a-) GaN, $SiO_2$, and diamond. Within this region, a $SiO_x$ region ($L_{a-SiOx}$) is sandwiched by a hybrid state known as the diffusion region, which extends a length of $L_{Diffusion}$ on each side.

To generate an amorphous structure in each region (left diffusion, $SiO_x$, and right diffusion regions), a step-by-step procedure for each separate region was followed. First, the region of interest was transitioned into an NPT ensemble for 20 ps, where it was heated to 5000 K to melt the crystal structure rapidly[49,50]. The melt state was maintained for 45 ps. Subsequently, the temperature of the hot areas was gradually reduced to 300 K over 200 ps, resulting in solidification[26,49,50]. Finally, the entire system was relaxed using NPT and NVT ensembles for 200 ps[51,52]. Throughout relaxation, a Nosé–Hoover thermostat was used to ensure temperature stability.

Periodic boundary conditions were used for the $y$- and $z$-directions, but a fixed boundary condition was used for the $x$-direction. To establish a steady temperature distribution, two Langevin thermostats ($T_H$ and $T_L$) were connected to the ends of the system, $T_H = 325$ K and $T_L = 275$ K, to evaluate TBR at 300 K. Once at steady state, the cumulative energy ($\Delta E$) added or subtracted to the heat source or sink region over a period of 1 ns was assessed. The energy change per unit time was obtained by linear fitting to the accumulated energy ($\Delta E$) data, which was then used to calculate the heat flux ($J = \Delta E/\Delta t$). TBR ($R$) was obtained as $R = (S \cdot \Delta T)/J$, where $S$ represents the cross-sectional area and $\Delta T$ is the temperature difference. For details of the typical heat flux calculation and temperature profile, see Fig. S10 in the SI.

**FEM Simulations**

Effects of interfacial thermal conductance on device temperature were investigated by solving the steady-state heat conduction equation using FEM in COMSOL Multiphysics 5.3a. For the boundary conditions, we set all surfaces except for the undersurface as 100 % thermal insulators. All other simulations parameters are listed in Table S2. Literature values of the thermal conductivity of diamond and TBR of the GaN/diamond interface are listed in Table S2. Note that for studies using the bonding method where the thermal conductivity of the diamond was not given, we used an identical value to the present study (approximately 2000 W/m·K).

**Acknowledgement**
This research was funded in part by JSPS KAKENHI (Grant Nos. 22H04950 and 22K14189). B.X. acknowledges support from the MST Foundation.

**Author contributions**
**Junichiro Shiomi**: Conceptualization; Supervision; Writing – Original Draft; Writing – Review & Editing; Discussion. **Bin Xu**: Methodology; Investigation; Data Curation; Writing – Original Draft; Visualization; Writing – Review & Editing; Discussion. **Fengwen Mu**: Methodology; Investigation; Discussion. **Rulei Guo**: Methodology; Discussion. **Shiqian Hu**: Methodology; Investigation; Software; Data Curation; Discussion. **Yingzhou Liu**: Methodology; Investigation; Software; Data Curation; Discussion.

# Figures

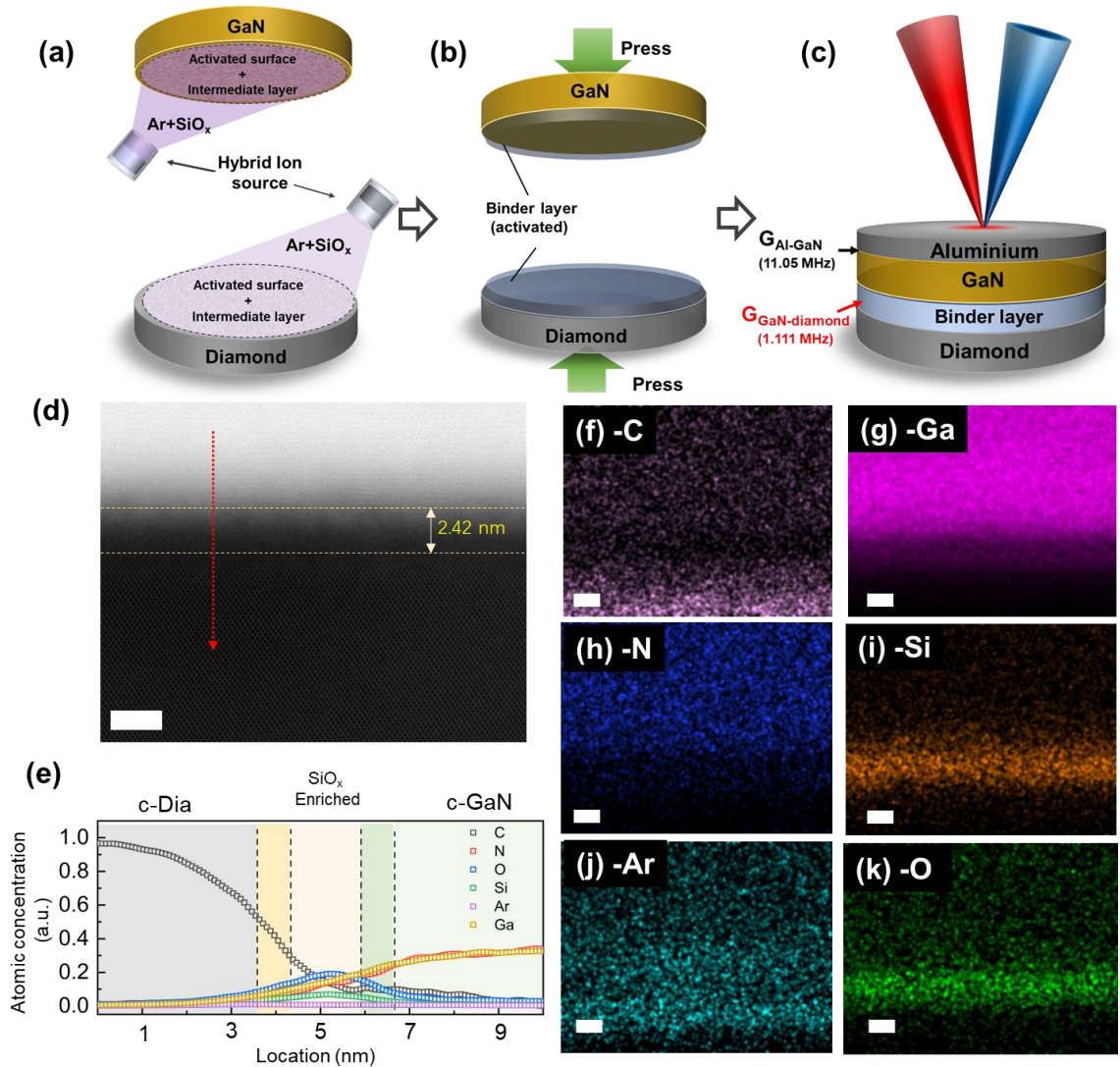

**Figure 1. Hybrid surface-activated bonding (SAB).** Schematic of (a),(b) SAB and (c) thermal boundary conductance measurement by TDTR. (d) STEM image and (e) corresponding EDS elemental line scan and (f)–(k) full region EDS mapping of #1; direction of linescan marked by a red arrow in (d); scale bars in the TEM image and EDS maps are 2 nm.

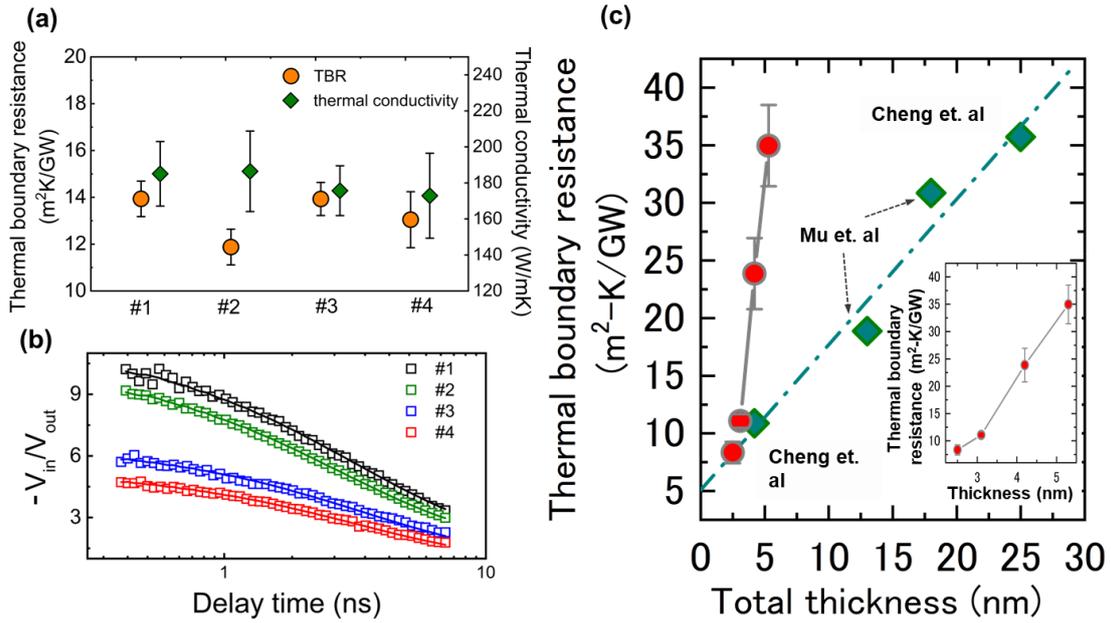

**Figure 2. TDTR measurement for GaN-on-diamond system.** (a) Thermal boundary resistance of Al/GaN interface (RAl-GaN) and the thermal conductivity of the GaN layer ($\kappa\_GaN$). (b) Temperature decay profile measured by TDTR (marks) and the theoretical fitting curves for samples #1–#4 at a modulation frequency of 1.111 MHz; (c) thermal boundary resistance of GaN/diamond interface (RGaN-diamond) as a function of the total interfacial thickness (red circle: present work, green diamond: previous reports[18,24]); inset is the result of present work plot to a smaller thickness range.

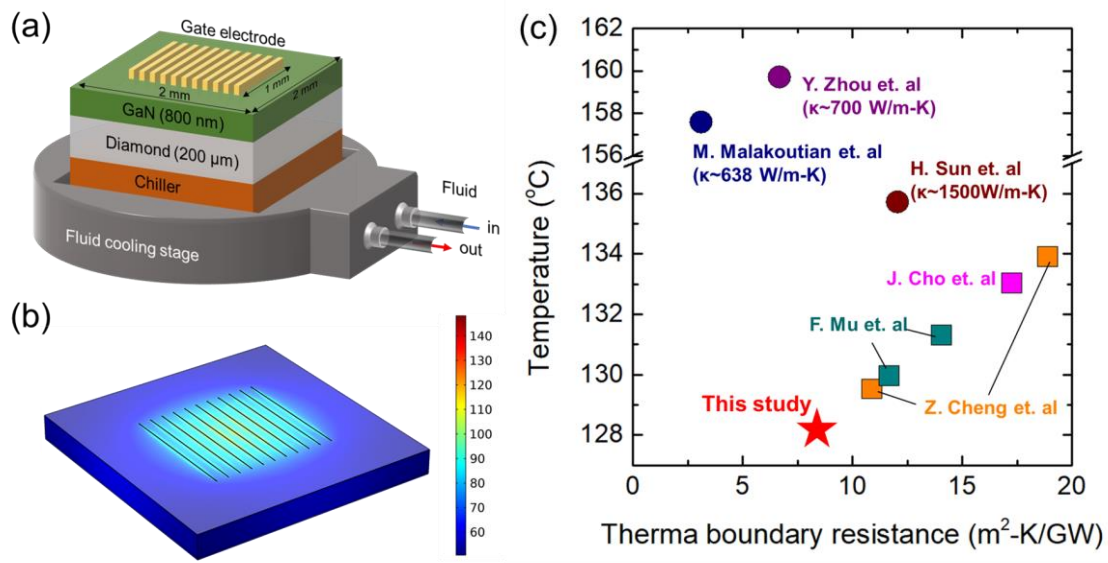

**Figure 3. Finite element method (FEM) simulation of GaN-on-diamond devices**. (a) Thermal boundary resistance of Al/GaN interface (RAl-GaN) and the thermal conductivity of the GaN layer ($\kappa_{GaN}$). (b) Temperature decay profile measured by TDTR (marks) and the theoretical fitting curves for samples #1–#4 at a modulation frequency of 1.111 MHz; (c) thermal boundary resistance of GaN/diamond interface (RGaN-diamond) as a function of the total interfacial thickness (red circle: present work, green diamond: previous reports[18,24]); inset is the result of present work plot to a smaller thickness range.

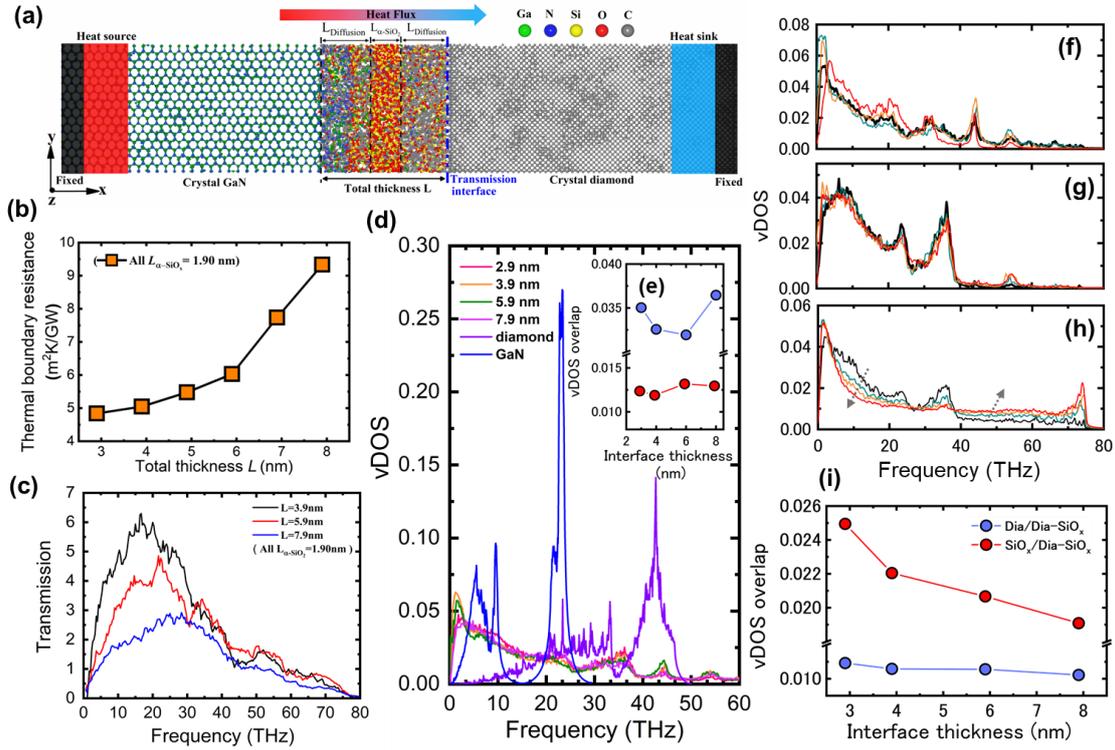

**Figure 4. Interfacial heat conduction calculated by MD simulation.** (a) Schematic of a typical MD simulations model with a layer-like amorphous interface. (b) TBR and (c) corresponding heat carrier transmission of GaN/diamond interface with different interfacial thickness; SiOx layer fixed to 1.9 nm. (d) vDOS of GaN, diamond, and the entire interfacial layer with a thickness of 2.9, 3.9, 5.9, and 7.9 nm. (e) vDOS overlap between GaN/interfacial layer (blue symbol) and diamond/interfacial layer (red symbol). vDOS of (f) GaN-SiOx diffusive region, (g) SiOx region, and (h) diamond-SiOx diffusive region. The interfacial layer thicknesses are 2.9 (black), 3.9 (green), 5.9 (orange), and 7.9 nm (red); grey arrow shows the trend in vDOS at individual frequencies ranges. (i) vDOS overlap between SiOx/diamond-SiOx diffusion region and diamond/diamond-SiOx diffusion region, respectively.

**Table**

Table 1. Thickness of interfacial layer of sample #1- #4.

| Sample No. | Total interfacial layer thickness (nm) | Individual diffusive layers thickness (nm) | SiO$_x$ layer thickness (nm) |
|---|---|---|---|
| #1 | 2.5 | 0.3 | 1.9 |
| #2 | 3.1 | 0.6 | 1.9 |
| #3 | 4.2 | 1.1 | 2.0 |
| #4 | 5.3 | 1.1 | 3.1 |

# Supplementary Information

## Low thermal boundary resistance at bonded GaN/diamond interface by controlling ultrathin heterogeneous amorphous layer

Bin Xu[1,2#], Fengwen Mu[3#*], Yingzhou Liu[4], Rulei Guo[1], Shiqian Hu[4**], Junichiro Shiomi[1,2***]

[1] Department of Mechanical Engineering, The University of Tokyo, 7-3-1 Hongo, Bunkyo, Tokyo 113-8656, Japan.

[2] Institute of Engineering Innovation, The University of Tokyo, 2-11 Yayoi, Bunkyo, Tokyo 113-8656, Japan.

[3] Innovative Semiconductor Substrate Technology Co., Ltd. No.25, Huayuan North Road, Haidian District, Beijing 100083, China.

[4] School of Physics and Astronomy, Yunnan University, 650091 Kunming, China.

[#]These authors contribute equally

Corresponding author:

*mufengwen@isaber-s.com

**shiqian@ynu.edu.cn

***shiomi@photon.t.u-tokyo.ac.jp

**S1. Details of MD simulations**

**2.1 Tersoff potential**

The Tersoff potential function is a model used to describe the interaction between covalent atoms, taking into account the concept of bond order. It considers not only the distance between atoms but also the bonding direction, making it essential to consider the influence of surrounding atoms when calculating the interaction between covalently bonded atoms. As a result, the Tersoff potential typically comprises two distinct components: a two-body interaction part and a three-body interaction part. In the potential energy file provided by LAMMPS, the first three columns have specific roles. Element1 represents the central atom involved in the three-body interaction, element2 corresponds to the atom bonded with the central atom, and element3 denotes the atom that influences the 1-2 bonds in terms of bond order. The parameters *n*, *beta*, *lambda2*, *B*, *lambda1* and *A* are exclusively used to describe the two-body interactions. On the other hand, the parameters *m*, *gamma*, *lambda3*, *c*, *d* and *costheta0* specific to the three-body interactions. The variables *R* and *D* are utilized to describe both two-body and three-body interactions[45–48].

In the context of two-body interactions, the selection of parameters depends primarily on element 1, which represents the central atom. Consequently, based on the three established Tersoff potential fields used to describe crystalline diamond[43,44], GaN[45], and $SiO_2$[46], the following judgments can be made: I. When element 2 is different from element 3, the corresponding two-body interactions parameters n, beta, lambda2, B, lambda1, and A are set to 0; II When element 2 and element 3 are the same, for dimensionless parameters, their values are equal to the parameter values of element 1. For parameters with dimensions, arithmetic mixing rules are followed.

In the context of the three-body interaction, both elements 1 and 3 have a combined influence on the selection of parameters. It is worth noting that in the current implementation of the Tersoff potential in LAMMPS, the parameter 'm' is restricted to the values 3 or 1. When m is set to 3 and gamma is equal to 1, the Tersoff potential field conforms to the original Tersoff form[47,48]. Conversely, when m is set to 1, the potential formula follows the form proposed by Albe et al.[45] Based on the Tersoff potential fields for diamond and $SiO_2$, m is assigned the value 3, while for GaN, m is assigned the value 1. Therefore, taking the above analysis into consideration, the following judgments can be made regarding the parameters of the three-body potential: I. When elements1 and 3 are either N or Ga elements, m is set to 1. Otherwise, it is set to 3. II. When element 1 is either N or Ga, gamma takes the corresponding value. Otherwise, it is set to 1. III. When both element 1 and element 3 are either N or Ga, the value of lambda3 corresponds to the parameters of N or Ga, respectively. Otherwise, it is set to 0. Additionally, the parameters c, d, and costheta are associated with the modified attraction potential, primarily dependent on the central atom. Therefore, these parameters are set equal to the value of element 1[47].

In both the two-body and three-body interactions, the selection of the parameters $R$ and $D$ follows the reference[45,46,48]. The details are as follows: For the parameter $R$: If element 1 and element 3 are the same, the value of $R$ for that atom is taken directly. On the other hand, if element 1 and element 3 are different, the value of $R$ is calculated as the square root of the product of the $R$ values for element 1 and element 3, i.e., $R=(R_1*R_3)^{1/2}$. For the parameter $D$: If element 1 and element 3 are the same type of atom, the value of $D$ is the same as the corresponding $R$ value. However, if element 1 and element 3 are different, a new value of $D$ is calculated based on the $R$ and $D$ parameters for element 1 and element 3. The formula is as follows: $R=(((R_1+D_1)^{1/2}*(R_3+D_3)^{1/2}-(R_1*R_3)^{1/2}))/2$.

Based on the aforementioned calculation details, we are also providing a Tersoff potential field file named "GaNSiOC.tersoff" in the attachment. This file is specifically designed for performing calculations involving the Ga, N, Si, O, and C elements using the LAMMPS software.

## 2.2 Phonon transmission calculation

Phonon transmission, which represents the heat flow across a cross-section (A), can be determined through molecular dynamics calculations that consider all orders of anharmonic interactions in the simulation[53,54],

$$\Gamma(\omega) = \frac{q(\omega)}{k_B \Delta T}, \tag{S1}$$

where $\Gamma(\omega)$ is phonon transmission, $k_B$ is the Boltzmann constant, and $\Delta T$ is the temperature difference between two Langevin thermostats. $q(\omega)$ is the heat flow across the cross-section $A$,

$$q(\omega) = \frac{2}{AM\Delta t} Re \sum_{i \in L} \sum_{j \in R} \langle \hat{F}_{ij}(\omega) \hat{v}_i(\omega)^* \rangle, \tag{S2}$$

where $\Delta t$ is the time interval between samples taken in the simulation and $M$ is the number of samples. $\hat{F}_{ij}(\omega)$ and $\hat{v}_i(\omega)^*$ are the Fourier transformation of the total force and the velocity of the atom, respectively. The calculation of the phonon transmission involves retaining only the forces exerted by the left part of the atoms on the right part, considering that $L$ and $R$ represent the left and right ends of the virtual interface in the simulated structure.

## 2.3 The calculation of the phonon density of states

The phonon density of states, denotes as $D(\omega)$, can be obtained using the following expression:

$$D(\omega) = \lim_{\tau \to \infty} \int_{-\tau}^{\tau} \frac{\langle \sum_{i=1}^{n} v_i(t) \cdot v_i(0) \rangle}{\langle \sum_{i=1}^{n} v_i(0) \cdot v_i(0) \rangle} e^{-i2\pi ft} dt, \tag{S3}$$

where $v_i$ represents the velocity of the *i*th atom, *f* is the frequency, *n* is the total number of atoms, *t* is the correlation time[55].

**Table**

Table. S 1 The parameters used in FEM calculation

| | |
|---|---|
| **Device power** | 110 W |
| **Device (gate) length** | 1 m |
| **Device (gate) thickness** | 200 nm |
| **Device (gate) width** | 4 μm |
| **Thermal conductivity of Gate** | 200 W/m-K |
| **Number of device (Gate)** | 11 |
| **Interval between devices (Gates)** | 0.1 mm |
| **GaN layer thickness** | 800 nm |
| **GaN layer length** | 2 mm |
| **Thermal conductivity of GaN** | 150 W/m-K |
| **Diamod Layer thickness** | 0.2 mm |
| **Diamond layer length** | 2 mm |
| **Thermal conductivity of diamond** | 2000 W/m-K |
| **TBR between gate and GaN** | 14 m$^2$-K/GW |
| **Heat transfer coefficient (bottom surface)** | 65 KW/m$^2$-K |
| **Ambient/cooling water temperature** | 20.15 ºC |

Table S2 The parameters of previous reported GaN-diamond interfaces for FEM calculation.

| Reference | Method | $R_{GaN\text{-}diamond}$ (m²-K/GW) | $\kappa_{Dia}$ (W/m-K) |
|---|---|---|---|
| This study | SAB | 8.3 | 2000 |
| Z. Cheng et. al[18] | SAB | 10.9 | 2000 |
|  |  | 18.9 | 2000 |
| F. Mu et. al[24] | SAB | 14.0 | 2000 |
|  |  | 11.6 | 2000 |
| J. Cho et. al[25] | High Temperature bonding | 18.8 | 2000 |
| M. Malakoutian et al.[1] | CVD | 3.1 | 638 |
| Y. Zhou et al.[22] | CVD | 6.7 | 700 |
| H. Sun et al.[23] | CVD | 12.0 | 1500 |

Table S3 The parameters used in TDTR analysis.

| Layer | Thickness (nm) | Heat capacity (KJ/m³-K) | TBC (MW/m²-K) | Thermal conductivity (W/m-K) |
|---|---|---|---|---|
| Aluminum | 100 | 243 | 70 (Al/GaN) | 150 |
| GaN | 1800 | 264 | 30/120 (GaN/diamond) | 185 |
| Diamond | $5 \times 10^3$ | 180 | - | 2045 |

**Figure**

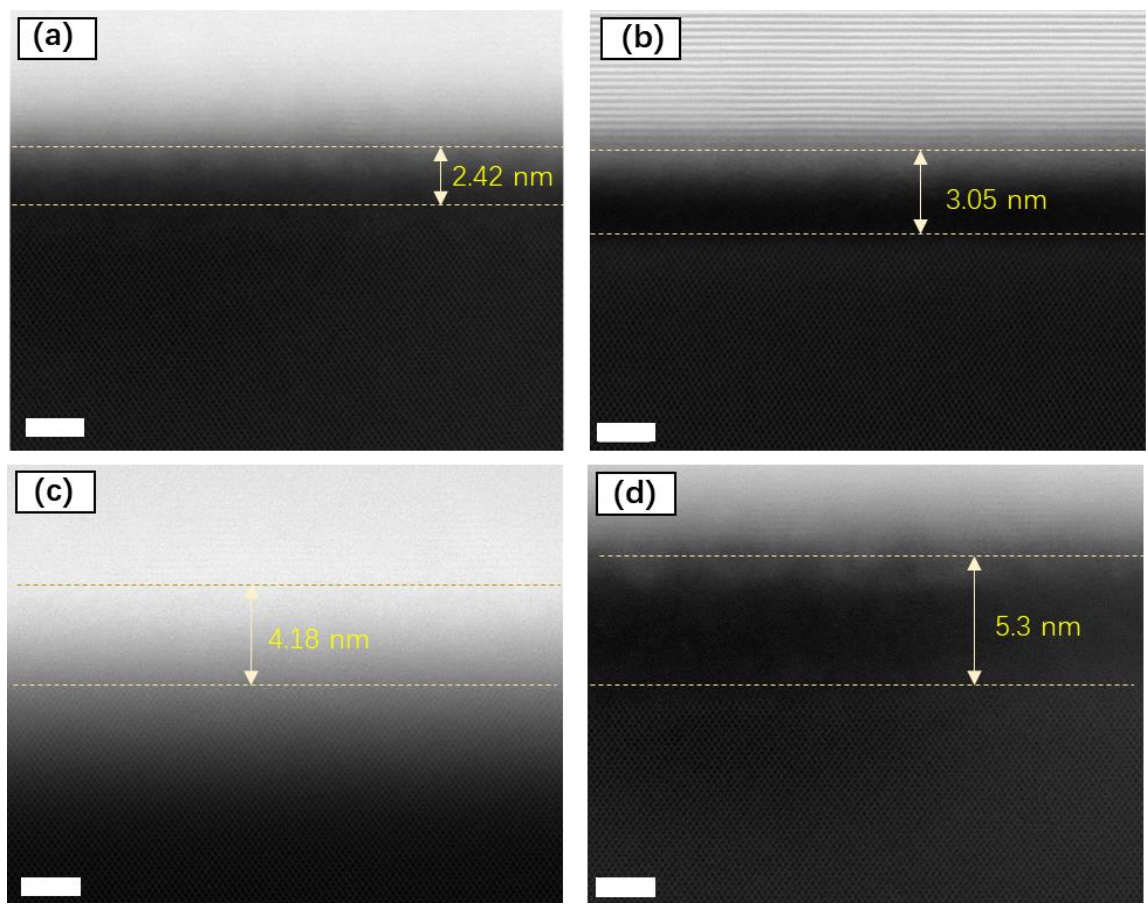

**Figure S1 STEM HADDF image of GaN/diamond interface.** (a-d) The cross-section view of sample #1~#4. The scale bar is 2 μm. The amorphous interfacial region is marked by yellow dash line.

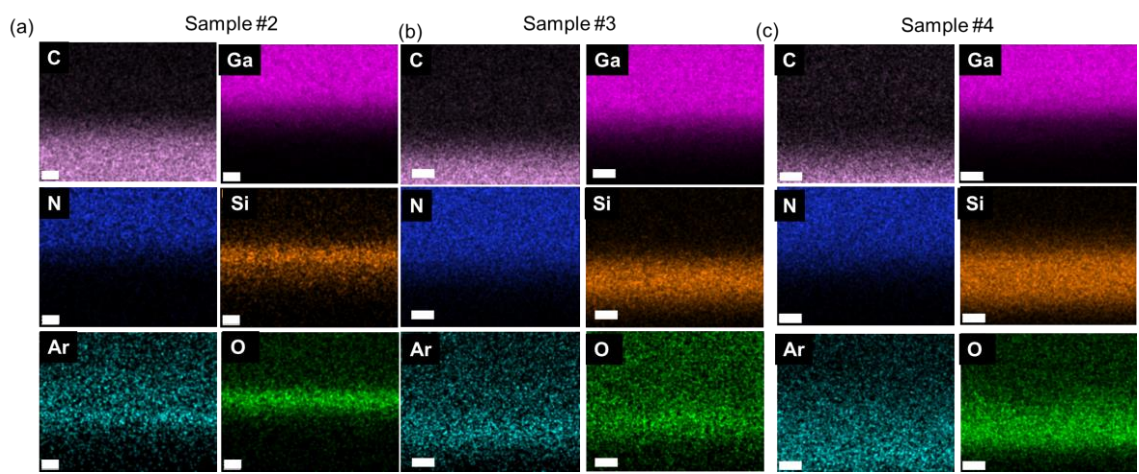

**Figure S2 EDS mapping of GaN/diamond interface.** (a-c) The elemental mapping of C, Ga, N, Si, Ar, O for sample #2~#4. The scale bar is 2 μm.

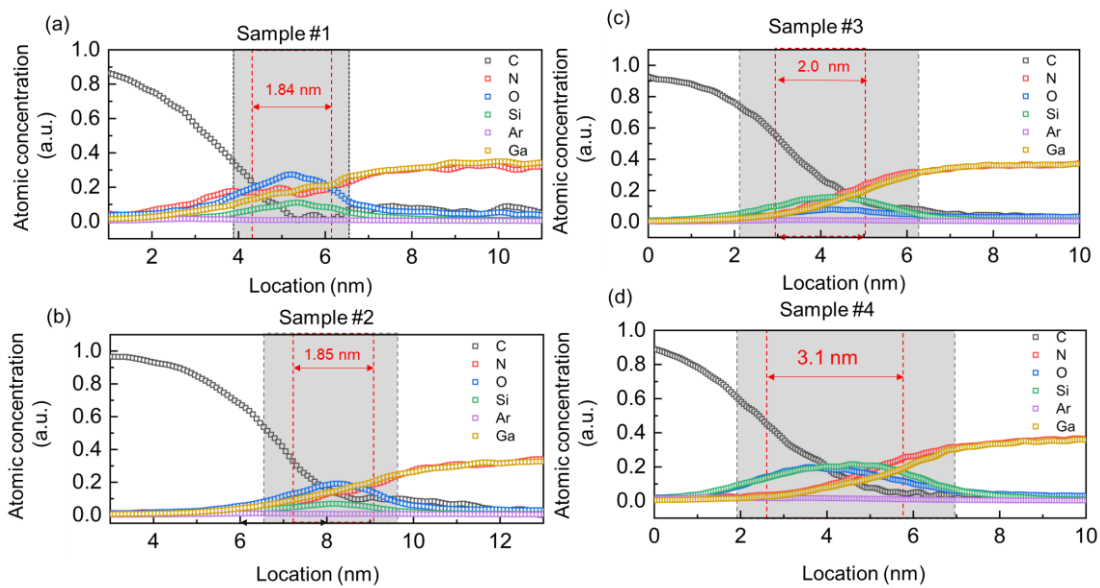

**Figure S3 EDS line scan of GaN/diamond interface.** (a-d) The elemental line scan of C, Ga, N, Si, Ar, O for sample #1~#4. The region in gray indicates the amorphous interfacial layer obtained from STEM image, and the red lines marked the Si-enriched

region where the normalized intensity is more than 1/2 of the maximum value.

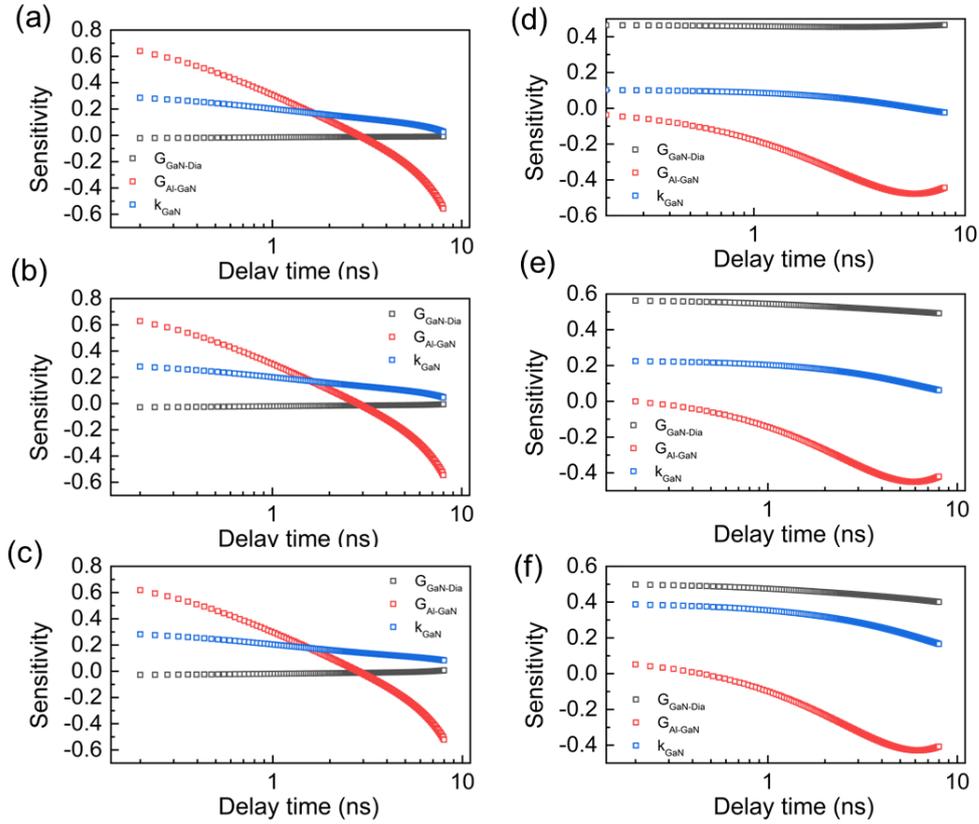

**Figure S4 TDTR sensitivity calculation. Sensitivity under** (a) 11.05 MHz, $G_{GaN-Dia}$=30 MW/m$^2$-K; (b) 11.05 MHz, $G_{GaN-Dia}$=60 MW/m$^2$-K, (c) 11.01 MHz, $G_{GaN-Dia}$=120 MW/m$^2$-K, (d) 1.111 MHz, $G_{GaN-Dia}$=30 MW/m$^2$-K, (e) 1.111 MHz, $G_{GaN-Dia}$=60 MW/m$^2$-K, (f) 1.111 MHz, $G_{GaN-Dia}$=120 MW/m$^2$-K.

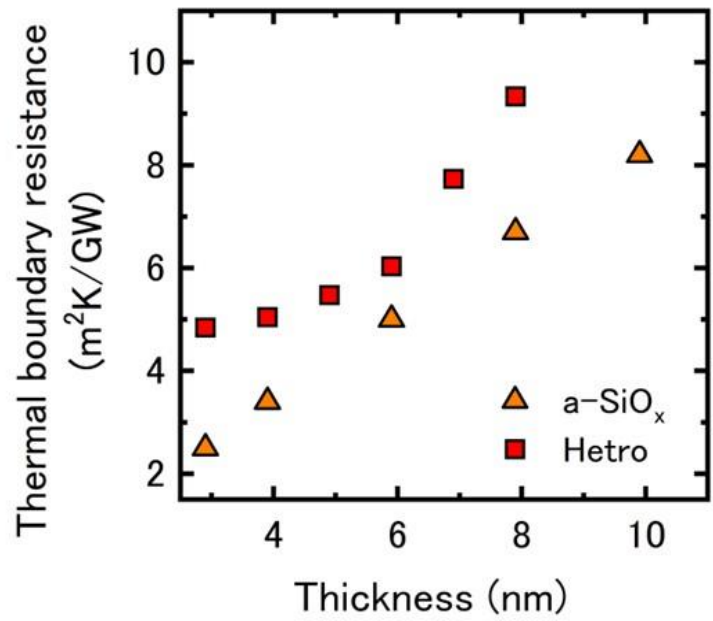

**Figure S5. MD simulations.** TBR of GaN/diamond interface with a heterogeneous layered interface and a homogeneous a-SiO2 interface.

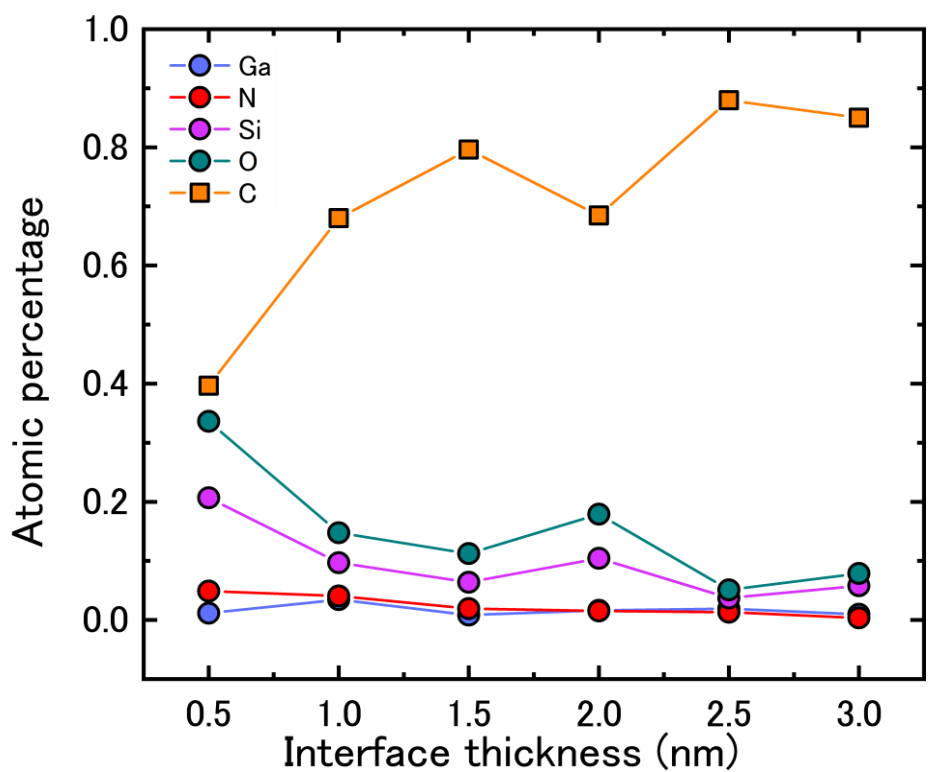

**Figure S6. MD simulations.** The concentration of individual elements in $SiO_x$/diamond diffusive layer with different thickness

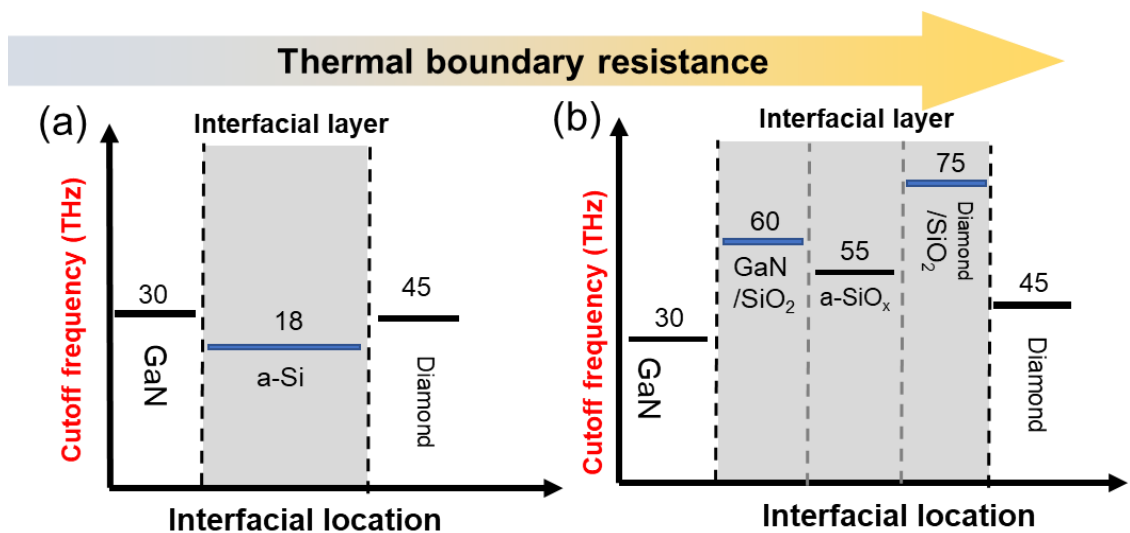

**Figure S7. Schematic of the distribution of cutoff frequency at the interface.** (a) The monolayer interface with a-Si, (b) The layer-like interfacial layer.

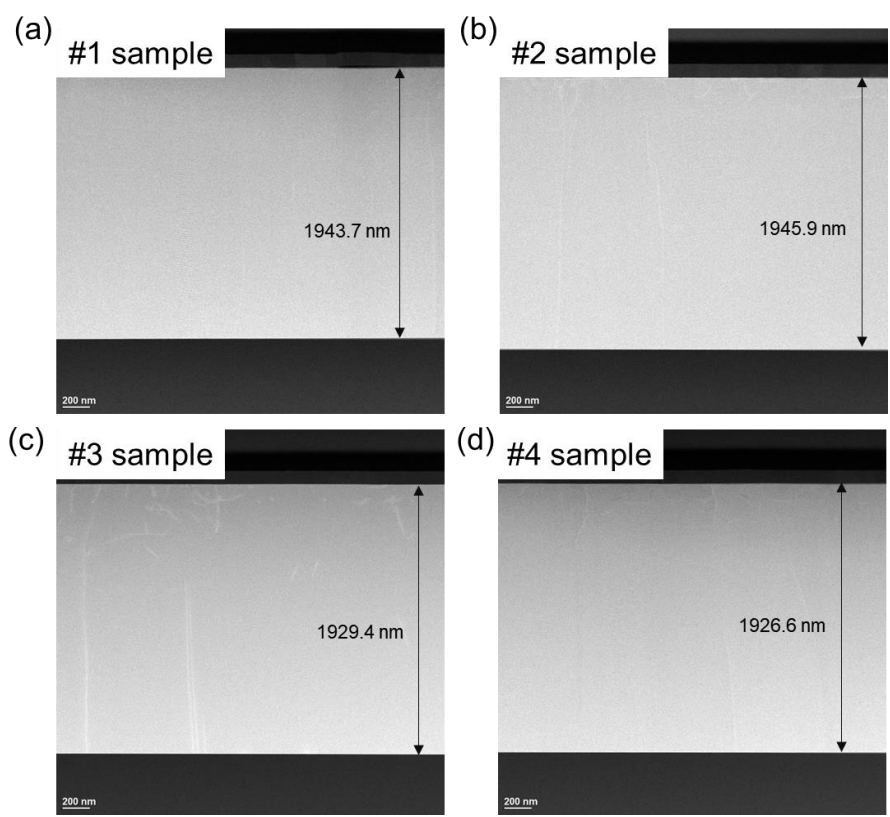

**Fig. S8. STEM image of GaN layer.** (a-d) The cross-section view (BF mode) of sample #1~#4.

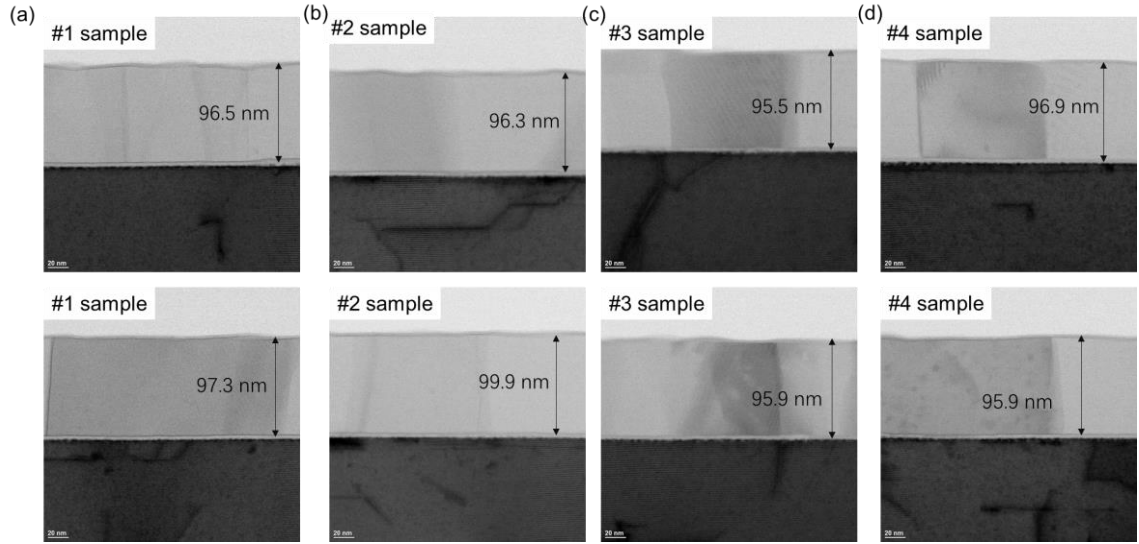

**Fig. S9. STEM image of Aluminum layer. (a-d)** The cross-section view (BF mode) of sample #1~#4. For individual samples, two location was observed.

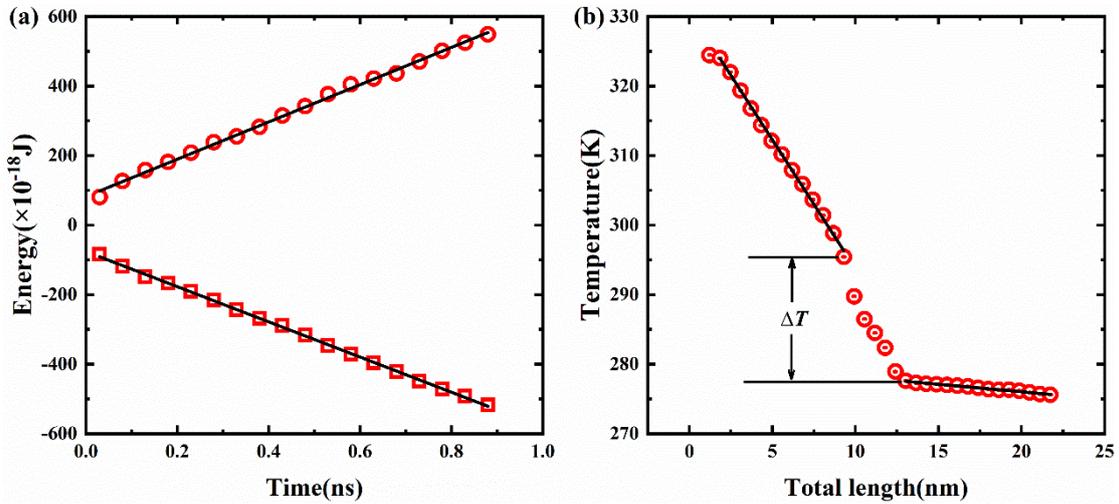

**Figure S10** (a) The typical cumulative energy change in the heat source and heat sink region, where the linear fitting is performed to obtain the energy change per unit time, which is used to calculate the heat flux. (b) The typical temperature profile, where the linear fitting is performed to obtain the difference of temperature.

# Low thermal boundary resistance at bonded GaN/diamond interface by controlling ultrathin heterogeneous amorphous layer

Bin Xu[1,2#], Fengwen Mu[3#*], Yingzhou Liu[3], Rulei Guo[1], Shiqian Hu[4*], Junichiro Shiomi[1,2*]

[1] Department of Mechanical Engineering, The University of Tokyo, 7-3-1 Hongo, Bunkyo, Tokyo 113-8656, Japan.

[2] Institute of Engineering Innovation, The University of Tokyo, 2-11 Yayoi, Bunkyo, Tokyo 113-8656, Japan.

[3] Innovative Semiconductor Substrate Technology Co., Ltd. No.25, Huayuan North Road, Haidian District, Beijing 100083, China.

[4] School of Physics and Astronomy, Yunnan University, 650091 Kunming, China.

#These authors contribute equally

*Corresponding author:

mufengwen@isaber-s.com

shiqian@ynu.edu.cn

shiomi@photon.t.u-tokyo.ac.jp

# S1. Details of MD simulations

## 1.1 Tersoff potential

The Tersoff potential function is a model used to describe the interaction between covalent atoms, considering the concept of bond order. It considers not only the distance between atoms but also the bonding direction, making it essential to consider the influence of surrounding atoms when calculating the interaction between covalently bonded atoms. As a result, the Tersoff potential typically comprises two distinct components: a two-body interaction part and a three-body interaction part. In the potential energy file provided by LAMMPS, the first three columns have specific roles. Element1 represents the central atom involved in the three-body interaction, element2 corresponds to the atom bonded with the central atom, and element3 denotes the atom that influences the 1-2 bonds in terms of bond order. The parameters *n*, *beta*, *lambda2*, *B*, *lambda1* and *A* are exclusively used to describe the two-body interactions. On the other hand, the parameters *m*, *gamma*, *lambda3*, *c*, *d* and *costheta0* specific to the three-body interactions. The variables *R* and *D* are utilized to describe both two-body and three-body interactions(*1–4*).

In the context of two-body interactions, the selection of parameters depends primarily on element 1, which represents the central atom. Consequently, based on the three established Tersoff potential fields used to describe crystalline diamond(*5*, *6*), GaN(*4*), and SiO$_2$(*3*), the following judgments can be made: I. When element 2 is different from element 3, the corresponding two-body interactions parameters n, beta, lambda2, B, lambda1, and A are set to 0; II When element 2 and element 3 are the same, for dimensionless parameters, their values are equal to the parameter values of element 1. For parameters with dimensions, arithmetic mixing rules are followed.

In the context of the three-body interaction, both elements 1 and 3 have a combined influence on the selection of parameters. It is worth noting that in the current implementation of the Tersoff potential in LAMMPS, the parameter 'm' is restricted to the values 3 or 1. When m is set to 3 and gamma is equal to 1, the Tersoff potential field conforms to the original Tersoff form(*1*, *2*). Conversely, when m is set to 1, the potential formula follows the form proposed by Albe et al.(*4*) Based on the Tersoff potential fields for diamond and SiO$_2$, m is assigned the value 3, while for GaN, m is assigned the value 1. Therefore, taking the above analysis into consideration, the following judgments can be made regarding the parameters of the three-body potential: I. When elements1 and 3 are either N or Ga elements, m is set to 1. Otherwise, it is set to 3. II. When element 1 is either N or Ga, gamma takes the corresponding value. Otherwise, it is set to 1. III. When both element 1 and element 3 are either N or Ga, the value of lambda3 corresponds to the parameters of N or Ga, respectively. Otherwise, it is set to 0. Additionally, the parameters c, d, and costheta are associated with the modified attraction potential, primarily dependent on the central atom. Therefore, these parameters are set equal to the value of element 1(*1*).

In both the two-body and three-body interactions, the selection of the parameters *R* and *D* follows the reference(*2–4*). The details are as follows: For the parameter *R*: If element 1 and element 3 are the same, the value of *R* for that atom is taken directly. On the other hand, if element 1 and element 3 are different, the value of *R* is calculated as the square root of the product of the *R* values for element 1 and element 3, i.e., *R=(R₁\*R₃)^(1/2)*. For the parameter *D*: If element 1 and element 3 are the same type of atom, the value of *D* is the same as the corresponding *R* value. However, if element 1 and element 3 are different, a new value of *D* is calculated based on the *R* and *D* parameters for element 1 and element 3. The formula is as follows: *R=(((R₁+D₁)^(1/2)\*(R₃+D₃)^(1/2)-( R₁\*R₃)^(1/2)))/2*.

Based on the aforementioned calculation details, we are also providing a Tersoff potential field file named "GaNSiOC.tersoff" in the attachment. This file is specifically designed for performing calculations involving the Ga, N, Si, O, and C elements using the LAMMPS software.

**1.2 Phonon transmission calculation**

Phonon transmission, which represents the heat flow across a cross-section (A), can be determined through molecular dynamics calculations that consider all orders of anharmonic interactions in the simulation(*7, 8*),

$$\Gamma(\omega) = \frac{q(\omega)}{k_B \Delta T}, \qquad (S1)$$

where $\Gamma(\omega)$ is phonon transmission, $k_B$ is the Boltzmann constant, and $\Delta T$ is the temperature difference between two Langevin thermostats. $q(\omega)$ is the heat flow across the cross-section $A$,

$$q(\omega) = \frac{2}{AM\Delta t} Re \sum_{i \in L} \sum_{j \in R} \langle \hat{F}_{ij}(\omega) \hat{v}_i(\omega)^* \rangle, \qquad (S2)$$

where $\Delta t$ is the time interval between samples taken in the simulation and $M$ is the number of samples. $\hat{F}_{ij}(\omega)$ and $\hat{v}_i(\omega)^*$ are the Fourier transformation of the total force and the velocity of the atom, respectively. The calculation of the phonon transmission involves retaining only the forces exerted by the left part of the atoms on the right part, considering that $L$ and $R$ represent the left and right ends of the virtual interface in the simulated structure.

**1.3 Calculation of phonon density of states**

The phonon density of states, denotes as $D(\omega)$, can be obtained using the following expression:

$$D(\omega) = \lim_{\tau \to \infty} \int_{-\tau}^{\tau} \frac{\langle \sum_{i=1}^{n} v_i(t) \cdot v_i(0) \rangle}{\langle \sum_{i=1}^{n} v_i(0) \cdot v_i(0) \rangle} e^{-i2\pi f t} dt, \qquad (S3)$$

where $v_i$ represents the velocity of the $i$th atom, $f$ is the frequency, $n$ is the total number of atoms, $t$ is the correlation time(9).

**Table**
Table. S 1 The parameters used in FEM calculation

|  |  |
|---|---|
| **Device power** | 110 W |
| **Device (gate) length** | 1 m |
| **Device (gate) thickness** | 200 nm |
| **Device (gate) width** | 4 μm |
| **Thermal conductivity of Gate** | 200 W/m-K |
| **Number of device (Gate)** | 11 |
| **Interval between devices (Gates)** | 0.1 mm |
| **GaN layer thickness** | 800 nm |
| **GaN layer length** | 2 mm |
| **Thermal conductivity of GaN** | 150 W/m-K |
| **Diamod Layer thickness** | 0.2 mm |
| **Diamond layer length** | 2 mm |
| **Thermal conductivity of diamond** | 2000 W/m-K |
| **TBR between gate and GaN** | 14 m$^2$-K/GW |
| **Heat transfer coefficient (bottom surface)** | 65 KW/m$^2$-K |
| **Ambient/cooling water temperature** | 20.15 °C |

Table S2 The parameters of previous reported GaN-diamond interfaces for FEM calculation.

| Reference | Method | $R_{GaN-diamond}$ (m²-K/GW) | $\kappa_{Dia}$ (W/m-K) |
|---|---|---|---|
| This study | SAB | 8.3 | 2000 |
| Z. Cheng et. al(10) | SAB | 10.9 | 2000 |
| | | 18.9 | 2000 |
| F. Mu et. al(11) | SAB | 14.0 | 2000 |
| | | 11.6 | 2000 |
| J. Cho et. al(12) | High Temperature bonding | 18.8 | 2000 |
| M. Malakoutian et al.(13) | CVD | 3.1 | 638 |
| Y. Zhou et al.(14) | CVD | 6.7 | 700 |
| H. Sun et al.(15) | CVD | 12.0 | 1500 |

Table S3 The parameters used in TDTR analysis.

| Layer | Thickness (nm) | Heat capacity (KJ/m³-K) | TBC (MW/m²-K) | Thermal conductivity (W/m-K) |
|---|---|---|---|---|
| Aluminum | 100 | 243 | 70 (Al/GaN) | 150 |
| GaN | 1800 | 264 | 30/120 (GaN/diamond) | 185 |
| Diamond | $5 \times 10^3$ | 180 | - | 2045 |

**Figure**

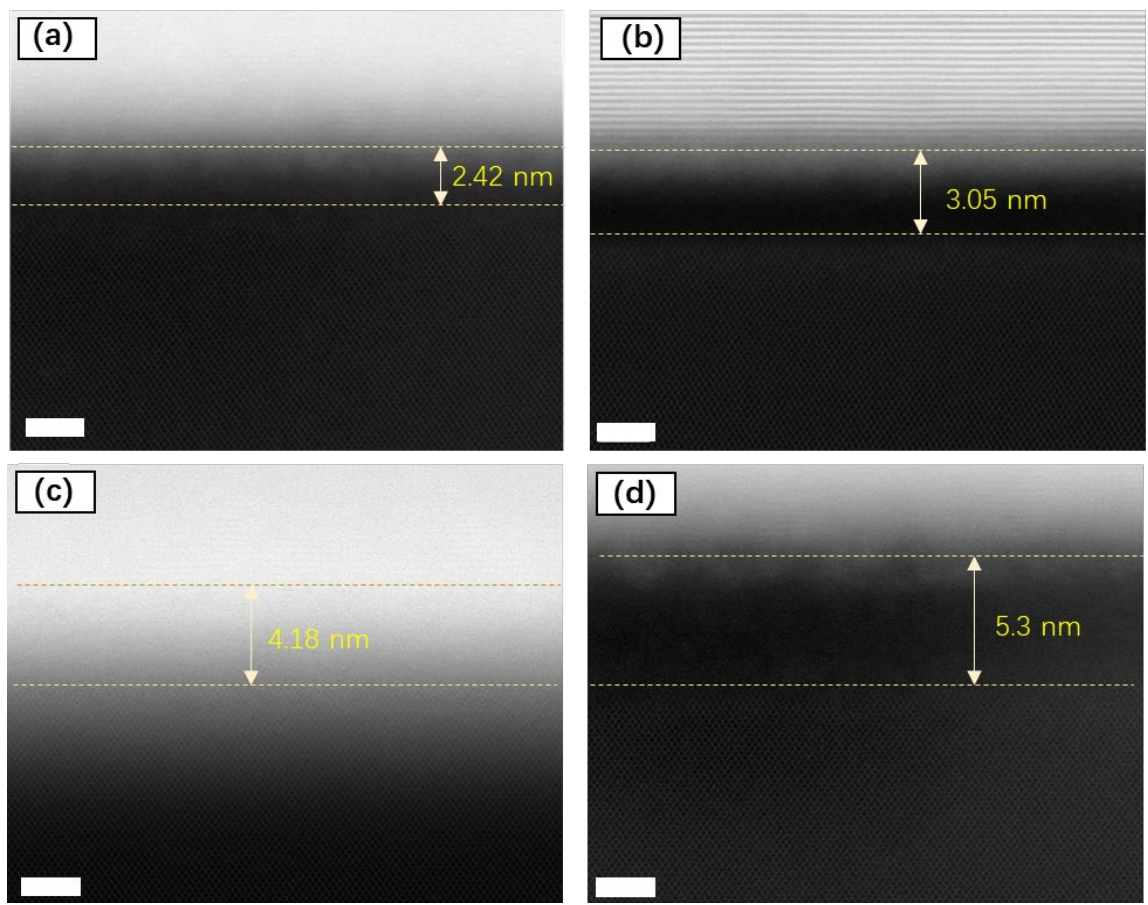

**Figure S1 STEM HADDF image of GaN/diamond interface.** (a-d) The cross-section view of sample #1~#4. The scale bar is 2 μm. The amorphous interfacial region is marked by yellow dash line.

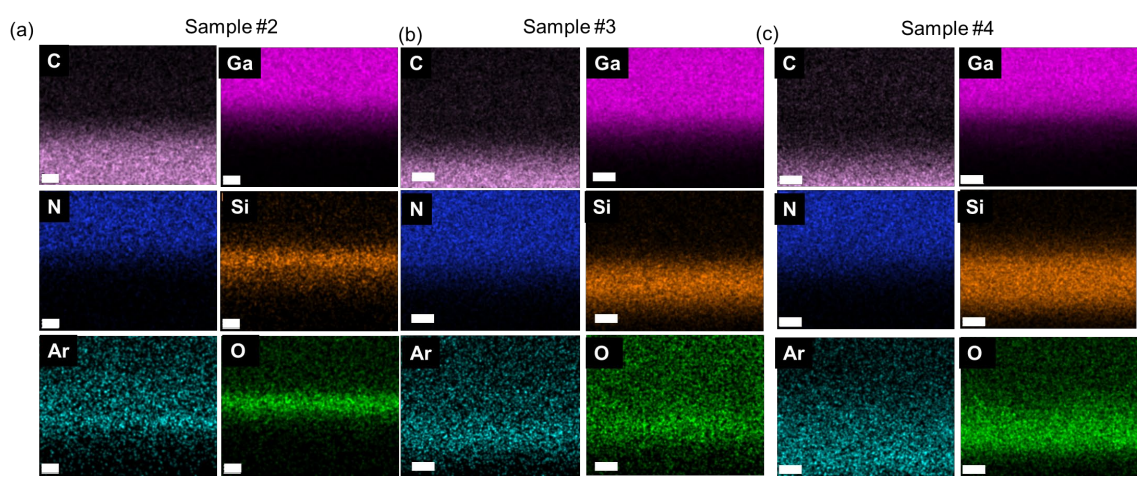

**Figure S2 EDS mapping of GaN/diamond interface.** (a-c) The elemental mapping of C, Ga, N, Si, Ar, O for sample #2~#4. The scale bar is 2 μm.

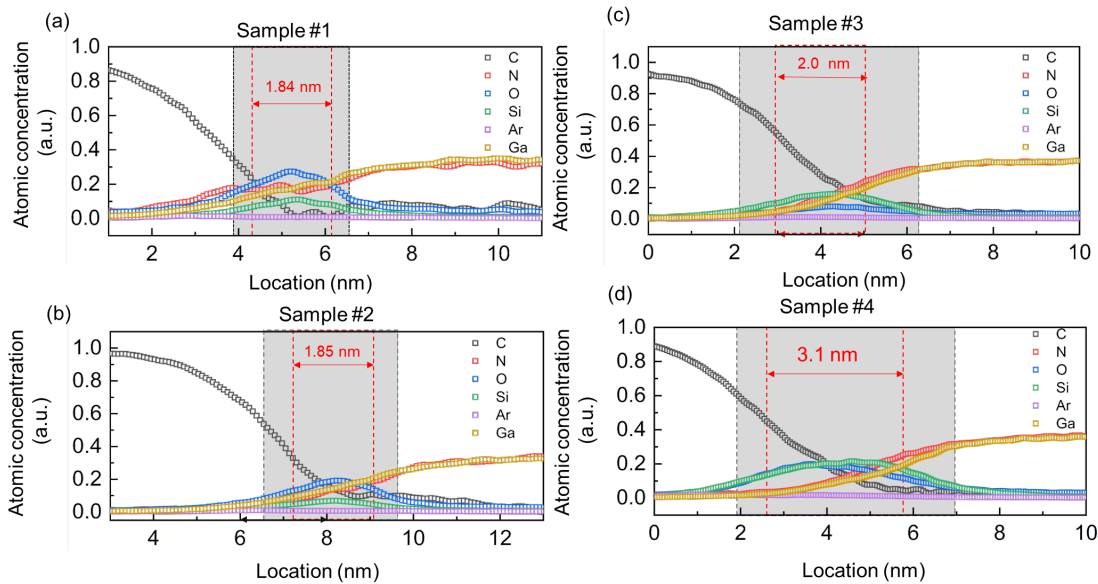

**Figure S3 EDS line scan of GaN/diamond interface.** (a-d) The elemental line scan of C, Ga, N, Si, Ar, O for sample #1~#4. The region in gray indicates the amorphous interfacial layer obtained from STEM image, and the red lines marked the Si-enriched region where the normalized intensity is more than 1/2 of the maximum value.

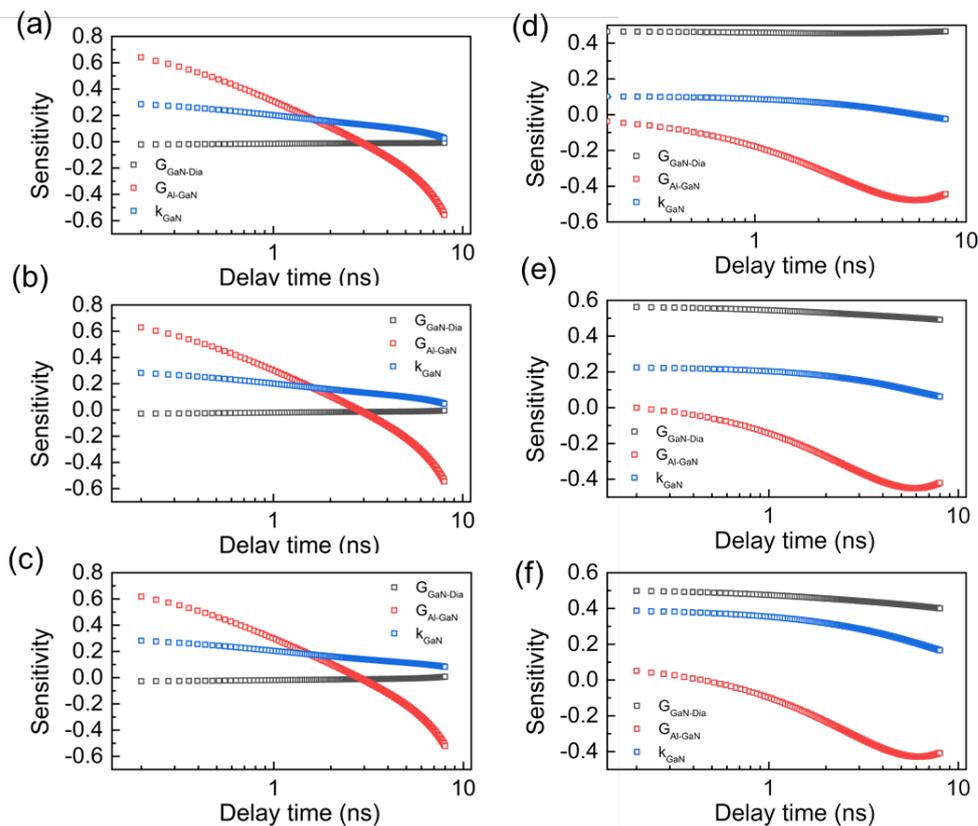

**Figure S4 TDTR sensitivity calculation.** Sensitivity under (a) 11.05 MHz, $G_{GaN-Dia}$=30 MW/m²-K; (b) 11.05 MHz, $G_{GaN-Dia}$=60 MW/m²-K, (c) 11.01 MHz, $G_{GaN-Dia}$=120 MW/m²-K, (d) 1.111 MHz, $G_{GaN-Dia}$=30 MW/m²-K, (e) 1.111 MHz, $G_{GaN-Dia}$=60MW/m²-K, (f) 1.111 MHz, $G_{GaN-Dia}$=120 MW/m²-K.

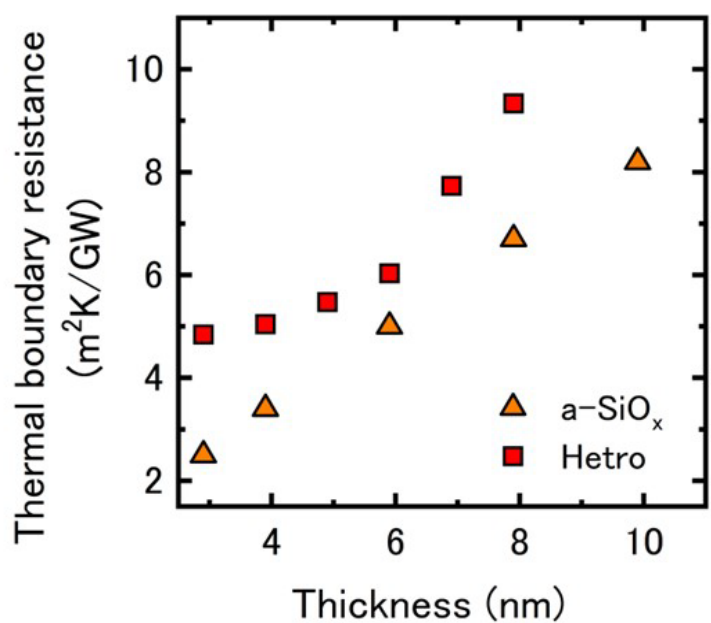

**Figure S5. MD simulations.** TBR of GaN/diamond interface with a heterogeneous layered interface and a homogeneous a-SiO2 interface.

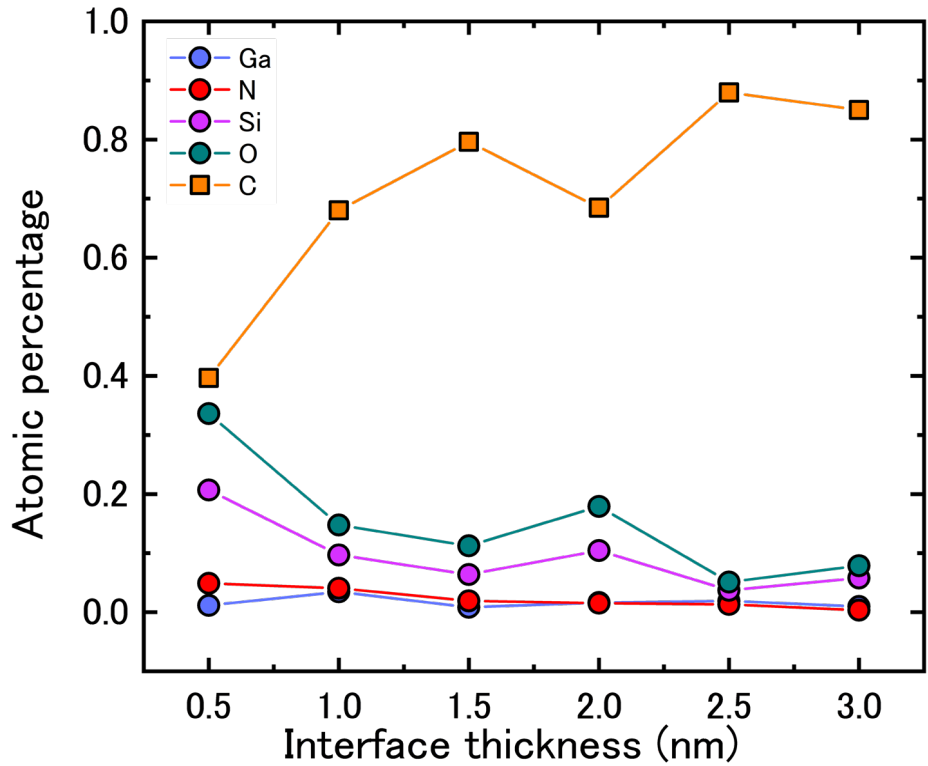

**Figure S6. MD simulations.** The concentration of individual elements in SiO$_x$/diamond diffusive layer with different thickness

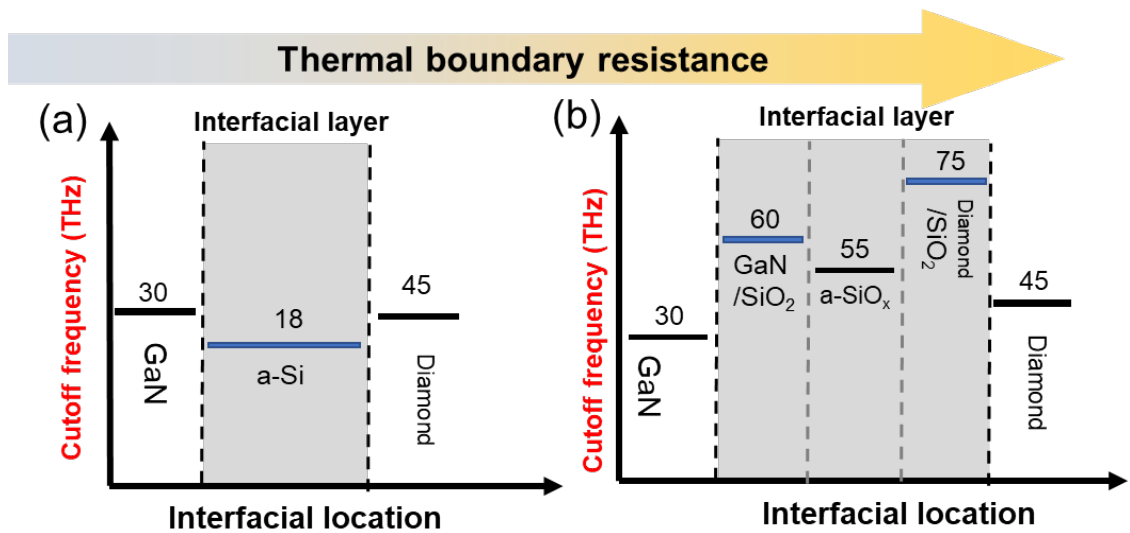

**Figure S7. Schematic of the distribution of cutoff frequency at the interface.** (a) The monolayer interface with a-Si, (b) The layer-like interfacial layer.

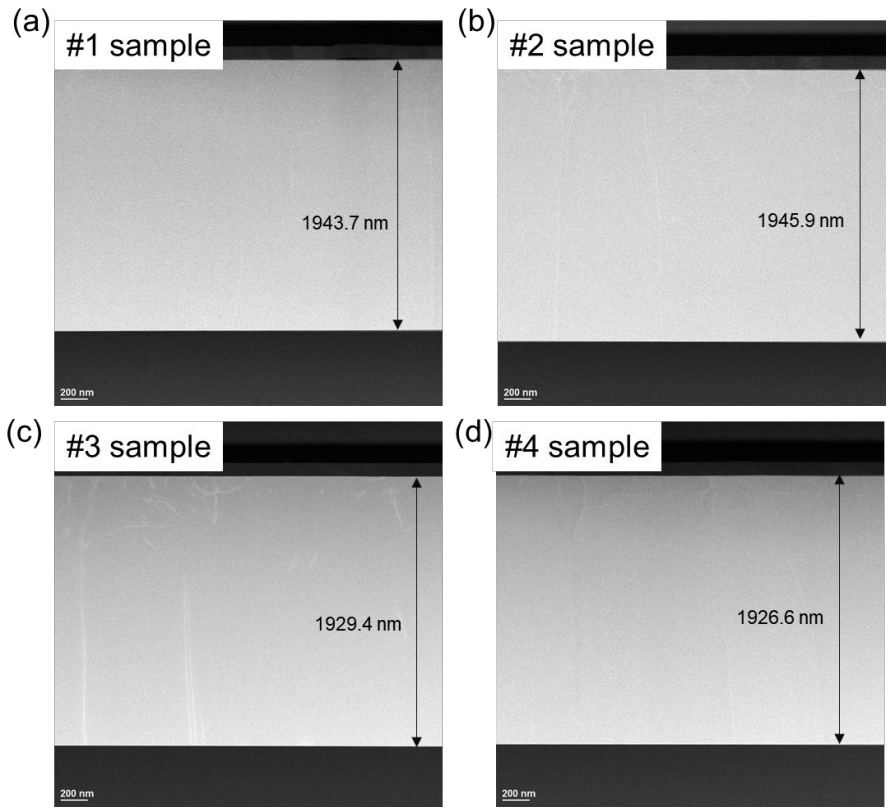

**Fig. S8. STEM image of GaN layer.** (a-d) The cross-section view (BF mode) of sample #1~#4.

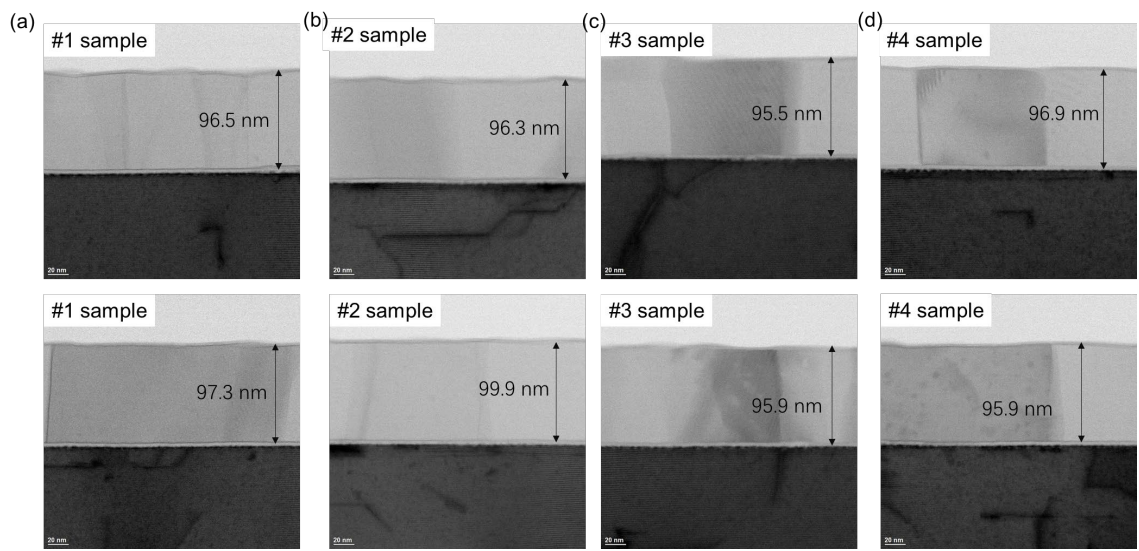

**Fig. S9. STEM image of Aluminum layer.** (a-d) The cross-section view (BF mode) of sample #1~#4. For individual samples, two location was observed.

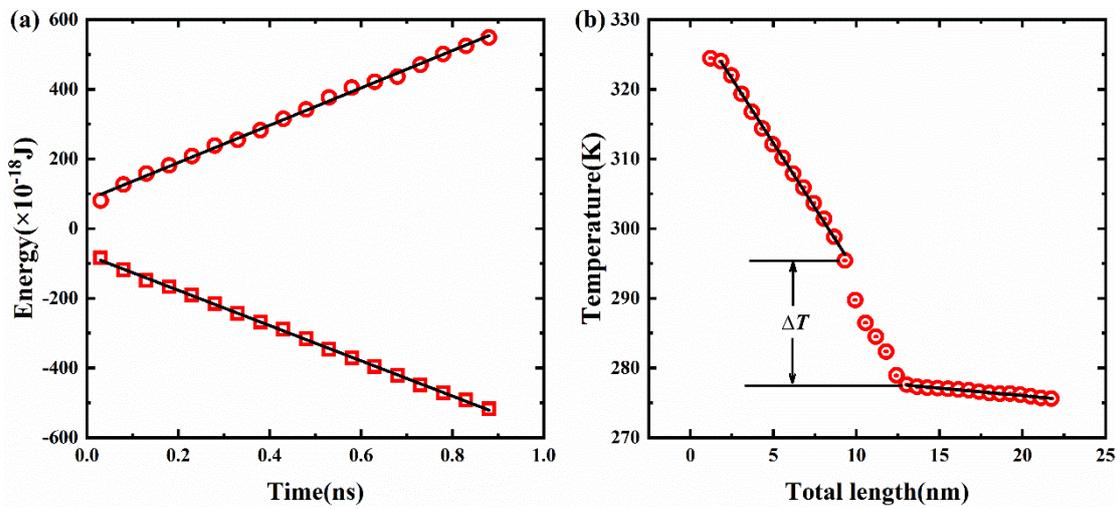

**Figure S10** (a) The typical cumulative energy change in the heat source and heat sink region, where the linear fitting is performed to obtain the energy change per unit time, which is used to calculate the heat flux. (b) The typical temperature profile, where the linear fitting is performed to obtain the difference of temperature.